\documentclass[fleqn,usenatbib]{mnras}

\usepackage{newtxtext,newtxmath}

\usepackage[T1]{fontenc}

\DeclareRobustCommand{\VAN}[3]{#2}
\let\VANthebibliography\thebibliography
\def\thebibliography{\DeclareRobustCommand{\VAN}[3]{##3}\VANthebibliography}

\usepackage{graphicx}	
\usepackage{amsmath}	
\usepackage{cleveref}
\usepackage[dvipsnames]{xcolor}
\definecolor{blue}{rgb}{0.26, 0.44, 0.8}
\definecolor{amaranth}{rgb}{0.9, 0.17, 0.31}
\definecolor{pink}{rgb}{0.57, 0.36, 0.51}
\definecolor{ao}{rgb}{0.0, 0.5, 0.0}
\definecolor{maroon}{rgb}{0.76, 0.13, 0.28}
\definecolor{cardinal}{rgb}{0.77, 0.12, 0.23}
\definecolor{frenchlila}{rgb}{0.53, 0.38, 0.56}
\usepackage{graphicx}  
\usepackage{color}
\hypersetup{pagebackref=false,
			hyperindex=false,
			citecolor=blue,
			colorlinks=true,
			linkcolor=blue,
			urlcolor=blue}

\DeclareMathOperator{\sech}{sech}

\title[\textit{GWs from DWDs as probes of the Milky Way}]{Gravitational Waves from Double White Dwarfs as probes of the Milky Way}

\author[Maria Georgousi et al.]{
Maria Georgousi$^{1,2}$\thanks{E-mail: \href{mailto:georgousi@iesl.forth.gr}{georgousi@iesl.forth.gr}},
Nikolaos Karnesis$^{1}$, Valeriya Korol$^{3,4}$, Mauro Pieroni$^{5}$ and Nikolaos Stergioulas$^{1}$
\\
$^{1}$Department of Physics, Aristotle University of Thessaloniki, Thessaloniki 54124, Greece\\
$^{2}$Institute of Electronic Structure and Laser, Foundation for Research and Technology-Hellas, PO Box 1527, 6R-71110 Heraklion, Greece\\
$^{3}$School of Physics and Astronomy \& Institute for Gravitational Wave Astronomy, University of Birmingham, Edgbaston, Birmingham B15 2TT, UK \\
$^{4}$Max-Planck-Institut f{\"u}r Astrophysik, Karl-Schwarzschild-Straße 1, 85741 Garching, Germany\\
$^{5}$Blackett Laboratory, Imperial College London, South Kensington Campus, London, SW7 2AZ,
UK
}

\date{Accepted XXX. Received YYY; in original form ZZZ}

\pubyear{2021}

\begin{document}
\label{firstpage}
\pagerange{\pageref{firstpage}--\pageref{lastpage}}
\maketitle

\begin{abstract}
Future gravitational wave detectors, such as the Laser Interferometer Space Antenna (\textit{LISA}), will be able to resolve a significant number of the ultra compact stellar-mass binaries in our own Galaxy and its neighborhood. These will be mostly double white dwarf (DWD) binaries, and their underlying population characteristics can be directly correlated to the different properties of the Galaxy. In particular, with \textit{LISA} we will be able to resolve $\sim\mathcal{O}(10^4)$ binaries, while the rest will generate a confusion foreground signal. Analogously to how the total electromagnetic radiation emitted by a galaxy can be related to the underlying total stellar mass, in this work we propose a framework to infer the same quantity by investigating the spectral shape and amplitude of the confusion foreground signal. For a fixed DWD evolution model, and thus a fixed binary fraction, we retrieve percentage-level relative errors on the total stellar mass, which improves for increasing values of the mass. At the same time, we find that variations in the Milky Way shape, at a fixed mass and at scale heights smaller than 500~pc, are not distinguishable based on the shape of stochastic signal alone. We perform this analysis on simulations of the LISA data, estimating the resolvable sources based on signal-to-noise criteria. 
Finally, we utilize the catalogue of resolvable sources to probe the characteristics of the underlying population of DWD binaries. We show that the  DWD frequency, coalescence time and chirp mass (up to $<0.7\,$M$_\odot$) distributions can be reconstructed from \textit{LISA} data with no bias.
\end{abstract}

\begin{keywords}
gravitational waves -- white dwarfs -- binaries:close -- Galaxy:structure 
\end{keywords}



%
%
%
%
\section{Introduction}

The Milky Way harbours a large variety of double compact objects composed of white dwarfs (WDs), neutron stars and black holes \citep[for a review see][]{amaro22}. Theoretical studies forecast hundreds of millions of double WDs (DWDs), and millions of double neutron stars and double black holes \citep[e.g.][]{Nelemans01a,rui10,yu10,nis12,lam18,bre20,vig20,wag21}. Although numerous, these are very challenging to detect through electromagnetic radiation because they either are too dim or do not emit light. The future Laser Interferometer Space Antenna (\textit{LISA}) \citep[][]{LISAwhitepaper} will survey these populations at shortest orbital periods delivering complete samples for periods of less than 17\,min within the Galaxy \citep{lam19,KorolHallakoun2021}. \textit{LISA} will measure the ensemble signal from these objects. However, out of millions, \textit{LISA} will be able to extract $\sim\mathcal{O}(10^4)$ DWDs, which makes them the most numerous type among \textit{LISA} sources \citep[e.g.][]{Korol2017,lam19,bre20,zen20,Wilhelm2021,thi21,tianquin}. The rest of compact Galactic binaries will blend together to form a confusion-limited foreground that is expected to affect the \textit{LISA} data at frequencies below a few mHz \citep[e.g.][]{LISAwhitepaper, Bender1997hs, nis12, rui10, Robson2017, KorolHallakoun2021}.

The Galactic confusion signal will have an overall spectral shape that depends on the properties of the actual DWD population. From earlier theoretical works, the spectrum is predicted to have two distinct attributes: the lower frequency tail ($\lesssim~\mathrm{mHz}$) and the higher frequency `knee' (at around few $\mathrm{mHz}$). Under the assumption of orbital evolution depending solely on gravitational-wave (GW) emission, the low-frequency tail is expected to follow a power-law with the index that should be $\sim 2/3$ in energy density $\Omega_\mathrm{gw}$ units (see for example~\cite{Phinney2001}). The form of the high-frequency `knee' structure depends on assumptions about the detectability of these sources with \textit{LISA}. For example, it heavily depends on the observation duration, and on the signal-to-noise (SNR) threshold criteria to classify the DWD sources as resolvable~\citep{Karnesis2021tsh, nis12, Timpano2006, crowder2007, Digman2022jmp}. Variations of intrinsic properties of the DWD population such as the chirp mass distribution can influence the detectability of binaries at few mHz, and, hence, the shape of the knee as well. For example, if DWDs are on average more massive than predicted by binary population synthesis studies and follow the observed mass function of single white dwarfs, the confusion foreground sharply drops at 2\,mHz instead of 3\,mHz \citep{KorolHallakoun2021}. This may seem like a small difference, however it bares important implications for the detectability of other \textit{LISA} sources, including merging massive black-hole binaries  \citep[$\sim10^{4} - 10^{7}$\,M$_\odot$, e.g.][]{Sesana2009,Klein2016,Dayal2019}, extreme-mass-ratio inspirals \citep[e.g.][]{Babak2017,moo17,bon20}, and backgrounds of cosmological origin \citep[e.g.][]{Caprini2016,Tamanini2016,Caprini2018mtu}. In addition, changes to initial binary fractions have been shown to affect the level of the Galactic confusion signal up to a factor of 2 \citep{thi21}. 

Several studies investigated the use of the resolved binaries for constraining the properties of the Milky Way \citep{adams2012,Korol2019,Wilhelm2021}. Only a few have exploited the confusion foreground to address similar constraints \citep{Benacquista2006, Breivik2020}. Using different strategies, both studies explored the effect of changing the disc scale height on the shape of confusion foreground only.
\citet{Benacquista2006} demonstrated that at a fixed
stellar space density, an increase in the scale height (leading to an increase in the total number of DWDs) raises the overall confusion level at frequencies lower than a few mHz and shifts the transfer frequency -- \emph{i.e.} the frequency above which \textit{LISA} can resolve all DWDs -- to slightly higher values. They also showed, which is  confirmed with our work here, that at a fixed total number of binaries, changing the scale heights does not influence the shape of the confusion foreground. On the other hand, in~\citet{Breivik2020}, the scale height was probed by decomposing the foreground signal on a basis of spherical harmonics. This methodology allows us to constrain the scale height of the Galaxy, even so with limited sensitivity compared to other methods, \emph{i.e.} in~\cite{adams2012} and \cite{Korol2019} using resolved DWDs. 

In this work, we fix a DWD population model and investigate different methods to characterize the properties of the Galaxy, by using two approaches. First, we describe the adopted fiducial population model in~\cref{sec:background}, and describe the methodology to estimate the residual foreground signal after subtracting the loudest sources in~\cref{sec:methodology}. Then, in~\cref{sec:Properties_impact}, as proposed in the earlier studies summarised above, we investigate the capabilities of \textit{LISA} to probe the scale height of the Galaxy ($Z_\mathrm{d}$), by measuring the confusion foreground signal. We do that by simulating  catalogues with different $Z_\mathrm{d}$, and estimating the confusion signal by evaluating the number of resolvable sources given the observation time of \textit{LISA}. By examining the SNRs and the waveform parameter errors of the recovered sources, we confirm that for small variations of $Z_\mathrm{d}$, it is challenging to measure the disc scale height from the confusion signal alone. We then go one step further to design a framework in order to assess the overall number of Galactic DWDs from the confusion signal, which can be linked to the total Galactic stellar mass. Interestingly, we find that number of sources can be reasonably well constrained by measuring the Galactic confusion noise. In~\cref{sec:hierarchical}, we perform the complementary analysis with the sub-set population of the resolvable sources. We use them to probe the probability distributions of three parameters of interest, namely the chirp mass, time of coalescence, and emission frequency. Finally, we draw our conclusions and dicuss our results in~\cref{sec:conclusions}.

This work is based on the Bsc thesis of~\cite{marythesis}.

%
%
%
%
\section{Mock double white dwarf population}
\label{sec:background}

To investigate the shape of the Galactic confusion noise we construct a fiducial Galactic DWD population based on the (publicly available) population synthesis code {\sc SeBa} \citep{PZ96,Nelemans01a,Toonen12}. The detailed description of this model is given in \citet{Toonen12}, to which we refer for further details. Below, we outline its most relevant features for this work. Our choice is motivated by the fact that this DWD evolution model yields the space density of DWDs in agreement with observations of the local white dwarf population and reproduces the general trend of the observed DWD mass ratio distribution \citep{Toonen12,Toonen2017}. We consider only one DWD evolution model and explore how the shape of the confusion noise changes when varying global properties of the Milky Way: its total stellar mass and shape (e.g. through the scale height parameter).

%
%
\subsection{Initial population}

The progenitor zero age main sequence population is assembled through a Monte Carlo technique. The mass of the primary stars -- the most massive in each pair -- is sampled from the initial mass function of \citet{KroupaIMF} between 0.95 - 10\,M$_\odot$, where the lower limit represents the minimum mass for an (isolated) star to reach the white dwarf stage in a Hubble time.  The mass of the secondary star is drawn uniformly between 0.08\,M$_\odot$ and the mass of the primary \citep{Raghavan2010,Duchene2013}. Initial binaries' semi-major axes are drawn from a log-uniform distribution extending up to $10^6\,$R$_\odot$ \citep{Abt1983,Raghavan2010,Duchene2013}. Orbit eccentricities are sampled from a thermal distribution \citep{Heggie1975}. The metallicity of the progenitor population is set to the Solar value, while the initial binary fraction is assumed to be of 50 per cent. Here we neglect a potential correlation between metallicity and initial binary fraction. Electromagnetic observations of close ($\lesssim$ 10\,au) low-mass binaries in the Solar neighborhood hint at a possible anti-correlation of binary fraction with metallicity \citep[][and references therein]{Bad18, Elb19, Moe19}. \citet{thi21} showed that, when implementing this anti-correlation in the binary population synthesis models for \textit{LISA}, the size of the population may decrease noticeably. We note that, in this work, we will neglect the degeneracy effects of the initial binary fraction to the total Galactic stellar mass (see~\cref{sec:Mass_impact}). However, in practice, the initial binary fraction represents a normalization factor and, if desired, the results can be re-scaled accordingly in the post-processing, or considered as a prior from electromagnetic observations~\citep{Duchene2013,Bad18,belokurov2020, Korol2020,maozhallakoun,napi}. 

%
%
\subsection{Binary evolution} 

Next, {\sc SeBa} evolves the progenitor population until both stars turn into white dwarfs following prescriptions for processes involved in the binary evolution, such as mass and  angular momentum transfer, common envelope evolution, magnetic braking, and gravitational radiation \citep[][and references therein]{PZ96, Toonen12}. In {\sc SeBa}, binaries starting on wide orbits ($ a \gtrsim$10\,au) end up evolving independently, as they would if they were born in isolation. However, to form a close binary system that falls in \textit{LISA}'s frequency window, at least one common envelope phase is required \citep{Paczynski1976,Webbink1984}. Specifically, to form a close DWD pair, the binary typically experiences at least two mass-transfer phases, of which at least one should be a common envelope \citep[e.g.][]{Nelemans01a}. 

Despite the importance of the common envelope evolution in the formation of double compact objects, there is no clear consensus regarding details governing this phase. Based on the observed DWDs,  \citet{Nelemans2000} determined the possible masses and radii of DWD progenitor stars, which were used to reconstruct binaries' past mass-transfer phases. Their work revealed that the DWD evolution models using the standard common-envelope formalism alone (\emph{i.e.} the $\alpha$-formalism, see \citealt{Ivanova2013} for details, equating the energy balance in the system and implicitly assuming angular momentum conservation) could not reproduce the observed mass ratios of DWDs. Therefore, they concluded that an alternative formalism for the first phase of the mass-transfer is required and proposed a formalism that parametrizes the angular momentum balance equation---instead of the energy balance equation---through a $\gamma$ parameter.
The result originally found by \citet{Nelemans2000} was later confirmed by studies using  larger DWD samples \citep{nel05,slu06} and by the binary population synthesis studies \citep{Nelemans01a, Toonen12}. 

We adopt a DWD evolution model based on the results prescribed above: a model combining the two aforementioned common envelope parametrizations, such that the first common envelope phase is typically described by the $\gamma$ formalism, while the second by the $\alpha$ formalism. In the standard {\sc SeBa} setting, $\gamma = 1.75$ and $\alpha \lambda =2$ (here $\lambda$ is a parameter dependent on the structure of the donor star) -- also obtained by fitting observations -- were assumed to generate our fidutial population. 

%
%
\subsection{The Milky Way model} \label{sec:mw}

To evaluate the total number of Galactic DWDs in the \textit{LISA}'s frequency window, and to assign their sky positions, distances, and present-day GW frequencies, a Galactic model is required. Practically, we set the stellar density distribution, Galactic age and star formation history. Our choice of the stellar density distribution affects DWDs' individual distances, while the choice of the Galactic age and SFH affect the present-day DWDs' GW frequencies; importantly, both also affect the total number ($N$) of DWD in the catalogue \citep[e.g.][]{Korol2021}.

Unlike binaries consisting of black holes and neutron stars, white dwarfs are born with no recoil kick, and, thus, DWDs are expected to follow the overall stellar density distribution. Although small recoil has been hinted in wide DWD systems \citep{El-Badry2018}, it should not significantly affect close binary systems detectable by \textit{LISA}. We consider a two-component density distribution composed of a central bulge and an extended (single-component) stellar disc as in \citet{Nelemans2004}. We note that the mass of the stellar halo is only $\sim 0.01$ of the disc stellar mass \citep[e.g.][]{bla16}, meaning that its contribution to the overall Galactic GW signal is negligible \citep[see also a dedicated BPS study by][]{Ruiter2009}. The two components are modeled as follows:
\begin{itemize}
    \item The stellar disc is decribed by an exponential radial stellar disc profile with an isothermal vertical distribution
        \begin{equation}  \label{eqn:disc}
        \rho_{\rm disc}(R,z) \propto (8\pi Z_{\rm d} R_{\rm d}^2)^{-1}  e^{-R/R_{\rm d}} \sech^2 \left( \frac{z}{Z_{\rm d}} \right),
        \end{equation} 
    where $0 \le R \le 19$ kpc is the cylindrical radius measured from the Galactic centre, $R_{\rm d} = 2.5\,$kpc is the characteristic scale radius, and  $Z_{\rm d}=300$ pc is the characteristic scale height of the disc \citep[e.g.][]{Juric2008,mac17}. The fiducial value of the disc scale height is chosen along the mean between the thin disc and the thick disc populations of red giants -- white dwarf progenitors -- for the typical age of LISA DWDs of several $\mathrm{Gyr}$~\citep[][see their figure 9]{mac17}. As in~\citet{Benacquista2006}, we also generate catalogues with $Z_{\rm d} = 100, 500\,$pc to assess the dependence of the shape of the Galactic confusion foreground with $Z_{\rm d}$. 
    \item The bulge component is modeled according to 
        \begin{equation} \label{eqn:bulge}
        \rho_{\rm bulge}(r) \propto (\sqrt{2\pi} r_{\rm b})^{-3} e^{-r^2/2 r_{\rm b}^2},
        \end{equation}
        where $r$ is the spherical distance from the Galactic centre, and $r_{\rm b}=0.5\,$kpc is the characteristic radius \citep[e.g.][]{Sofue2009}.
\end{itemize}
We set the Sun's position to (R$_\odot,z_\odot) = (8.1,0.03)~\mathrm{kpc}$~\citep[e.g.][]{abu19}.

To model the star formation history of the Galaxy, we use the plane-projected star formation rate from a chemo-spectrophotometric model of \citet{BP99} for the stellar disc, whereas to account for the bulge we double the star formation rate in the inner 3\,kpc.  We assume the age of the Galaxy to be of 13.5\,Gyr. The obtained result matches well the shape of the observed Galactic star formation history inferred from single white dwarf stars \citep[][]{Fantin2019}.  
The integral of the star formation rate over the time up to 13.5\,Gyr yields the total stellar mass of $\sim 8.2 \times 10^{10}\,$M$_\odot$, which breaks down to $2.6\times10^{10}\,$M$_\odot$ for the bulge component and $5.6\times10^{10}\,$M$_\odot$ for the disc component. 
We also verified that a constant star formation history, equivalent to our fiducial star formation history over the last several Gyr, have no significant effect on the shape of the Galactic GW foreground, and, thus, are not considered here as an alternative model. However, we expect that a significantly different star formation history (e.g. star formation rate that increases towards present time) may produce noticeable differences in the shape of Galactic GW foreground.

%
%

%
%
%
%
\section{Methodology}
\label{sec:methodology}

In this section, we briefly describe the methodology we followed in order to estimate the residual foreground signal, as well as the empirical spectral model that we used in order to characterize it for different population simulations.

\begin{figure*}
 	\includegraphics[width=.9\linewidth]{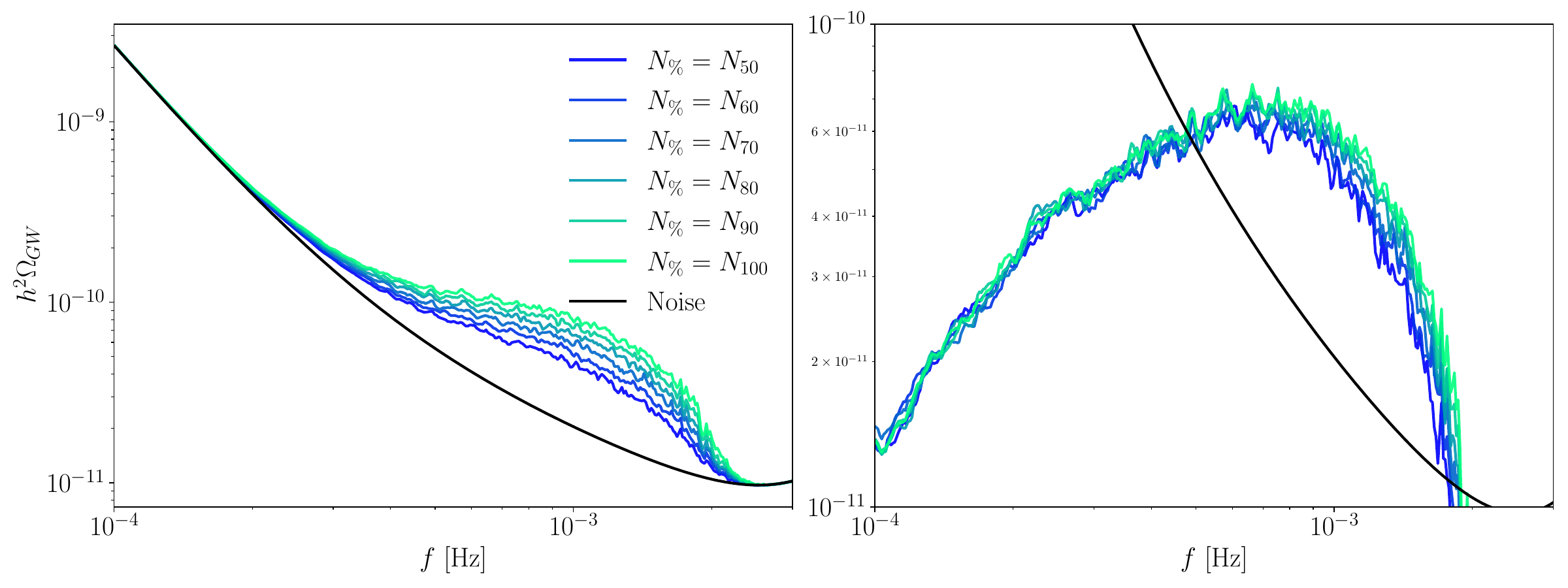}
 	\caption{\label{fig:mass_dep} \emph{Left}: Data (signal$+$noise) obtained after performing the procedure described in~\cref{sec:foreground_estimation} on the simulated datasets. The subscript denotes the percentage of binary sources of the galaxy ($N_\%$) in the dataset, while $N_{100}$ means the default value for the given population model. \emph{Right}: The signal components (same datasets as in the left panel, without the instrumental noise) re-scaled with $1 / N_\%$, compared with \textit{LISA} noise.}
\end{figure*}
 
%
%
\subsection{Estimation of foreground signal}
\label{sec:foreground_estimation}

It is expected that \textit{LISA} is going to be a signal-dominated observatory~\citep{LISAwhitepaper}, which means that millions of sources are going to be overlapping in the time and in the frequency domains. The most abundant type of GWs source is by far the ultra-compact stellar binaries, which are mostly comprised by DWDs. Resolving such overlapping signals, where their total number is also unknown, requires the employment of sophisticated stochastic algorithms. These data analysis methodologies are based on trans-dimensional Markov Chain Monte Carlo (MCMC) methods~\citep{rjmcmc}, where the source parameters and their actual number are being estimated simultaneously from the data. Such implementations have been already successfully demonstrated for the \textit{LISA} data by~\cite{corn07} and \cite{Litten2020}, or~\cite{Boileau2021}. However, these algorithms require  significant computational resources and computational time in order to achieve convergence. Usually, the analysis is being performed in short frequency segments (since the signal of interest is well localized in frequency), and the main result is then derived by the combination and post-processing of the analysis of the individual segments. Instead, we have chosen to work with an alternative method of estimating the foreground GW signal described in~\cite{Karnesis2021tsh, nis12, Timpano2006, crowder2007, Digman2022jmp}, which is based on SNR criteria. While this methodology follows simplified assumptions about data analysis with \textit{LISA}, it can be proven to be producing optimistic estimates of the extracted sources and confusion signal. We explain this methodology in detail below. 

We begin by generating the GW signals of each source in our simulated catalogue directly in the frequency domain. The waveform model is the one introduced in~\cite{corn07}, which is described by a given set of parameters $\vec{\theta} = \{ f_0, \dot{f}, \beta, \lambda, {\cal A}, \iota, \psi, \phi_0 \}$. The ${\cal A}$ corresponds to the strain amplitude, while $f_0$ is the main emission frequency and $\dot{f}$ its derivative. The $\psi$ characterizes the polarization and $\phi_0$ the initial phase of the wave, and finally the  $\beta$, $\lambda$, and $\iota$ are localization parameters. At this stage, we also compute the optimal SNR $\rho_\mathrm{opt}$ of each source in isolation, \emph{i.e.} when no other GW sources are present in the frequency bin. At this stage, $\rho_\mathrm{opt}$ is computed with respect to the instrumental noise only, and without taking into account other possibly overlapping signals. We then make a first estimate of the overall confusion signal $S_\mathrm{k}$, by computing the smoothed Power Spectral Density (PSD) directly on the simulated data. This is performed via a running median on the PSD of the data, and is then further processed by fitting a polynomial model or by applying a Gaussian kernel. Having an estimate of the confusion signal plus instrumental noise $S_\mathrm{k}$, we then compute the SNR $\rho_i$ of each source $i$ of the catalogue with respect to $S_\mathrm{k}$. If $\rho_i > \rho_\mathrm{thres}$, where $\rho_\mathrm{thres}$ a set SNR threshold, then the source $i$ is considered resolvable and thus subtracted from the data. Sources with very low $\rho_\mathrm{opt}$ are skipped in this calculation, and thus increase the efficiency of this methodology. After iterating once over the catalogue, the confusion signal $S_\mathrm{k+1}$ is estimated anew on the residuals of the first iteration. The procedure continues until there are no more sources to subtract, or if the residuals change negligibly with respect to the previous iteration. At the end of this process, we perform a Fisher Matrix (FIM) analysis on the resolvable sources, computed with respect to the final $S_\mathrm{final}$. The FIM is written as 
\begin{equation}
	\left. F_{ij} = \left( \frac{\partial h(\vec{\theta})}{\partial \theta_i} \bigg| \frac{\partial h(\vec{\theta})}{\partial \theta_j}  \right) \right\vert_{\vec{\theta} = \vec{\theta}_\mathrm{true}}^{S= S_\mathrm{final}},
	\label{eq:fim}
\end{equation}
where the $(\cdot|\cdot)$ denotes the noise weighted inner product between two time series $a$ and $b$:
\begin{equation}
	\left( a | b \right) = 2 \int\limits_0^\infty \mathrm{d}f \left[ \tilde{a}^\ast(f) \tilde{b}(f) + \tilde{a}(f) \tilde{b}^\ast(f) \right]/S(f)\,,
	 \label{eq:inprod}
\end{equation}
and $S$ the power spectral density of the noise.

In this work we have assumed the nominal mission duration which is $\mathrm{T}_\mathrm{obs} = 4~\mathrm{years}$, and we have adopted an SNR threshold of $\rho_\mathrm{thres} = 5$, which is used to classify the DWD binaries as detectable or not. One could adopt the more conservative limit of $\rho_\mathrm{thres}=7$, but we choose to work on and present the more optimistic case scenario following~\cite{crowder2007, Korol2020} and~\cite{blaut}. In essence, the main results of this paper would not be greatly affected, other than a small shift on the final model parameters (see below in~\cref{sec:Mass_impact}). Additionally, have also adopted the nominal `sciRD' sensitivity levels, which are explained in detail in~\cite{SciRD} and in~\cite{Babak2021mhe}. The data are simulated and analyzed directly in the noise orthogonal A and E Time Delay Interferometry (TDI) variables~\citep[][]{tdi, aet}.

Finally, we should also mention that this technique is based on a few ideal assumptions about the data. For example, being based solely on SNR criteria for assessing the detectability of a particular source, we ignore effects like signal overlap. At the same time, when a source is characterized as resolvable, we subtract it from the data at {\it true parameter values}, which means that we are left with perfect residuals. We also ignore any possible noise non-stationarities, such as gaps, glitches, or other time-variations of the noise PSD. In a slightly more realistic scenario, we would have to assume longer observation duration, {\it i.e. } $\mathrm{T}_\mathrm{obs} = 6~\mathrm{years}$ with a given duty cycle~\citep{missionduration}, and additionally take into account the presence of data gaps in our analysis. This would introduce complications in our simplified scheme here, because we would need to develop an analysis methodology for the treatment of the gaps, such as masking~\citep{dey21}, or gap-filling~\citep{Baghi19}. Nevertheless, for the case of monochromatic sources such as the DWD binaries in our work here, the SNR is proportional to the $\sqrt{\mathrm{T}_\mathrm{obs}}$, thus our simplifying approximation of simulating $\mathrm{T}_\mathrm{obs} = 4~\mathrm{years}$ of uninterrupted data is valid at first order~\citep{Korol2020, missionduration}. On the other hand, observing data with $\mathrm{T}_\mathrm{obs} = 6~\mathrm{years}$, even with data interruptions, would yield improved estimates on the $\dot{f}$ parameter of the waveform, which describes the evolution of the main emission frequency. This in turn would have a positive impact on the estimation of the chirp masses of the binaries (see~\cref{sec:hierarchical} for details), and thus our approximation is more conservative in this regard. In the end, being careful about the limitations mentioned above, this method is proven to be extremely fast, and even if it yields more optimal results than more statistically robust methods, it is a very useful tool to analyze large datasets, such as the DWDs population in this work.

At this point, we should comment to the differences to earlier works. According to~\cite{lam19}, the number of sources with well-determined chirp mass is around 5\,per cent after $4$ years of observations. However, we should point out that there are considerable differences between their simulations and those in this present work. The percentage of sources with measurable chirp mass is greatly dependent on the population model under study, and the given adopted SNR threshold for the analysis. In particular, for our population model, with a LISA mission duration of $T_\mathrm{obs} = 4$~$\mathrm{years}$, we recover around $25000$ binaries with SNR threshold $\rho_\mathrm{thres}=7$ and around $40000$ binaries with $\rho_\mathrm{thres}=5$ \citep[e.g.][]{Karnesis2021tsh}, while \cite{lam19} recover $12000$ binaries in total. In addition, with the lower $\rho_\mathrm{thres}=5$, we also recover a stochastic confusion noise signal with lower spectral amplitude, which also contributes to estimating better the parameters of the individual waveforms (thus, in some cases getting a better estimate for the chirp mass of the binaries). In this regard, we estimate that around $\sim15$\,per cent of the resolved sources have a relative error-bar for the chirp mass that is smaller than 10\,per cent, bringing their total number to $6000$ binaries. We also illustrate this point at the top right panel of figure~\ref{fig:zfisher}, where the results of the FIM are summarized for the resolved binaries of the particular population that we studied. Finally, it is also fair to point out that both our study and that of \citet{lam19} are based on the binary population synthesis technique. When forecasting DWD population detectable by LISA starting from the know DWD spectroscopic sample \citep{maozhallakoun}, \citet{KorolHallakoun2021} estimate even larger fraction of $\sim40$\,per cent (10000 DWDs) of resolved sources with a relative error-bar for the chirp mass that smaller than 10\,per cent (see their figure 4, and also section~\ref{sec:hierarchical}).

%
%
\subsection{Analytic model}
\label{sec:analytic_model}

In terms of the single-sided spectral density $S(f)$ of the gravitational-wave strain, we define the model for the stochastic component of the signal due to unresolved Galactic binaries, as~\cite{Karnesis2021tsh}:
\begin{equation}
	S_\mathrm{gal} (f) = \frac{A}{2}  f^{-n_s^S} e^{-(f/f_1)^\alpha} \left \{ 1 + \mathrm{tanh}\left[ \left( f_\mathrm{knee} - f \right) /f_2 \right] \right\} ,
	\label{eq:galfit_strain}
\end{equation} 
where $A$ is the amplitude of the signal, $n_s^S$ is the low-frequency spectral tilt\footnote{Notice the superscript $S$ to stress strain units.}, while the exponential term (with the two parameters $f_1$ and $\alpha$) models the `loss of stochasticity' due to the smaller density of sources at higher frequencies. Finally, the $\tanh$ term (with the two parameters $f_\mathrm{knee}$ and $f_2$) represents a signal cut-off due to individual removal of bright sources. While this model is mostly phenomenological, it gives a good insight on the effect of its parameters on the properties of the Galaxy. The spectral density $S(f)$ is related to the ratio $\Omega_{\rm gw}\equiv \rho_{\rm gw}/ \rho_c$, where $\rho_{\rm gw}$ is the energy density of gravitational waves, $\rho_{\rm c}=3c^2H_0^2/8\pi G$ and  $H_0 $ is the present value of the Hubble expansion rate, via 
\begin{equation}
	h^2 \, \Omega_{gw} \equiv \frac{4 \pi^2 f^3}{3 (H_0 /h)^2 } S(f) \; ,
\end{equation}
where $h$ is the normalized Hubble expansion rate, defined through $H_0 = h \times 100 \mathrm{~km} \mathrm{~s}^{-1} \mathrm{Mpc}^{-1}$. In the following, for the sake of analysis convenience, we cast the template in~\cref{eq:galfit_strain} as:
\begin{equation}
\begin{aligned}
	h^2 \, \Omega_\mathrm{gw, gal } & = 10^{ \log_{10}{( h^2 \, \Omega_\mathrm{gw}^*)}} \left( \frac{f}{f_*} \right)^{n_s} e^{-(f/10^{\log_{10} f_1})^\alpha} \\ 
	& \hspace{0.5cm} \times \left \{ 1 + \mathrm{tanh}\left[ \left( 10^{\log_{10} f_\mathrm{knee} } - f \right) / 10^{\log_{10} f_2 } \right] \right\} ,
	\label{eq:galfit}
\end{aligned}
\end{equation}
where $f_*$ is a fiducial pivot (for convenience in the following we set this as $f_* = 10^{-3}$Hz). Notice that all the parameters modeling amplitudes and frequencies have been expressed in terms of $\log_{10}$, in order for their numerical values to be of order unity. We have also included~\cref{fig:model} in the Appendix section, where the impact of the different terms appearing in~\cref{eq:galfit} is explored visually.

After executing the procedure described in~\cref{sec:foreground_estimation} we are left with an estimate of the residual foreground signal. Since in~\cref{sec:Properties_impact} of this work we are not interested in assessing the precision at which we will be able to estimate the parameters of each individual source, but rather in the impact of the galaxy's foreground signal to the underlying population model, we proceed without considering the instrument noise component. Noise will only be added at face value for plotting reasons (for example in the left panel of~\cref{sec:Mass_impact}), in order to place our analysis in context. 
 
\subsection{Data pre-processing \label{sec:preproc}}

Independently of the properties of the noise component, to reduce the numerical complexity of the problem, we proceed by using the techniques of~\cite{Caprini:2019pxz, Pieroni:2020rob, Flauger:2020qyi}. In these works, the effective mission duration of 3\,yr (\emph{i.e.} assuming 4\,yr with $75\%$ duty-cycle efficiency) is chopped into $N_c$ data segments having the same duration of around $11.5$ days, corresponding to a frequency resolution of $\simeq 10^{-6}$Hz. A Gaussian realization of both the signal and noise components is then generated for each of these segments and the spectrum at each frequency is then estimated to be the average over these $N_c$ realizations. On the other hand, in this work, and consistently with the procedure employed in~\cite{Karnesis2021tsh}, we perform some operations on the dataset provided by the procedure of~\cref{sec:foreground_estimation}, which accounts for the complete mission duration (meaning that in the frequency domain the frequency resolution is ${\sim} 1/\mathrm{T}_\mathrm{obs}$), to transform it into a dataset with the same statistical properties of the ones of~\cite{Caprini:2019pxz, Pieroni:2020rob, Flauger:2020qyi}. For this purpose, averaging over segments is equivalent to averaging over neighboring frequencies, \emph{i.e.} for a given frequency, say $f_i$, this corresponds to averaging over data points in the range $f_i-5\times 10^{-7}$Hz $\leq f \leq f_i+5\times 10^{-7}$Hz. At this point, in order to further lower the computational complexity of the problem, we coarse-grain (\emph{i.e.} bin) the data to a lesser dense set, by performing an inverse variance weighting of the data using the noise as an estimate for the variance (see sec. 3.1 of~\cite{Flauger:2020qyi} for details). In particular, for each decade in frequency above $10^{-4}$Hz we pass from the initial $10^{-6}$Hz linear spacing to $250$ evenly logarithmically spaced frequency points. Finally the instrument's response function is factored out in order to get the dataset, which can directly be compared with the model of~\cref{eq:galfit}. In the following sections, these datasets will be denoted as $D^{k}_{ij}$ where the two indexes $i,j \in \{ {\rm A, E, T}\}$ run over the different TDI combinations and the index $k$ runs over the different frequency coefficients of each dataset. The corresponding frequencies and weights are denoted with $f^{k}_{ij}$ and $w^{k}_{ij}$ respectively.

%
%
%
%
\section{Investigating the Galaxy properties from the measurement of the stochastic signal}
\label{sec:Properties_impact}

In this section we describe the different parameters we can characterize by analyzing the spectral shape of the confusion noise. We first proceed to design a framework to identify the overall Galactic stellar mass, and then focus on probing the scale height $Z_\mathrm{d}$, as done in the previous studies, but now using our developed framework.

%
%
 \subsection{Probing the total stellar mass of the Galaxy}
 \label{sec:Mass_impact}

\subsubsection{Concept}

It is intuitive to imagine that the total energy emitted in GWs by the Galactic DWD population is related to the total stellar mass of the Galaxy. This is analogous to how the total light emitted by a galaxy is set by its stellar mass. Indeed, the (stellar mass) of a galaxy is often determined from its measured luminosity via the stellar-mass-light ratio that can be derived from stellar population synthesis models \citep[e.g.][]{bru83,ren86, mar98,BP99,Nesti_2013}. These models combine stellar isochrones, spectral libraries and the initial stellar mass function with a star formation history to provide the stellar-mass-light ratio ---or even the entire spectral energy distribution of a galaxy---at a given time in a specific electromagnetic band. By fitting to synthetic stellar-mass-light ratios or to observations, one can recover the stellar mass of the galaxy. Along the same lines, here we explore how the foreground signal due to unresolved DWD sources measured by \textit{LISA} is related to the total stellar mass of the Galaxy. In particular, and for the first time, we attempt to connect the level (amplitude) of the measured stochastic confusion signal, parametrized with \cref{eq:galfit_strain}, with the total mass of the galaxy through the estimate of the total number of DWD binaries contributing to the foreground. 

When forward modeling \textit{LISA} observations, our assumptions regarding the Galactic model, the initial binary population and subsequent evolution until the DWD stage (cf. section \ref{sec:background}) set the number of detectable (as resolved and as a foreground) DWDs. Among these assumptions, the total mass of the Galaxy is found to have the strongest impact on the detectable sources in both space-based and ground-based GW detectors contexts \citep[e.g.][Keim et al., in prep.]{art2018,Korol2020,roe20}. Just like described above for electromagnetic observations, binary population synthesis models can be reverse engineered to determine the mass of the population based on detected sources \citep{Korol2021}. 

To explore how for a set DWD evolution model (cf. section \ref{sec:background}) the shape of the foreground signal changes with the Milky Way stellar mass, we generate $6$ different datasets, corresponding to different values of $m_{\mathrm{gal}}$, which translates to different number of binaries $N_{\mathrm{gal}}$. We achieve this by simply creating subsets of the same synthetic catalogue by randomly removing entries, according to the given final desired percentage of sources. In practice, we define $m_{\mathrm{G}} \equiv 8.2 \times 10^{10}\,$M$_\odot$, which corresponds to a particular number of total binaries $N_\mathrm{G} = N_{100}$. Then, for each catalogue $i$, we set $N_{\mathrm{i}} \equiv  N_{\%} N_{\mathrm{G}}$ with $N_{\%} = \{ 0.5, 0.6, 0.7, 0.8, 0.9, 1\}$. 

We then proceed to simulate the actual dataset for each of the $N_{\%}$ considered. We do that by directly computing the waveforms of each catalogue entry for the given mission duration ($\mathrm{T}_\mathrm{obs} = 4~\mathrm{years}$). We then estimate the confusion foreground signal due to the given DWD binary population, following the methodology presented in~\cref{sec:foreground_estimation}. From this part of the analysis, we recover the sub-catalogue of the resolvable sources, as well as an estimate of the residual confusion signal.

\subsubsection{Fitting the spectral shape}
\label{sec:fitting_the_shape}

In order to model the impact of the number of sources $N_{\%}$ on the spectral shape of the residuals foreground signal, we make use of the spectral model we had defined in~\cref{sec:analytic_model}, and in particular in~\cref{eq:galfit_strain}. Inspecting~\cref{eq:galfit_strain} in combination with the information from~\cref{fig:mass_dep}, we can design an analytic formulation that would account for the relation of $N_{\%}$ to the overall amplitude, as well as a slight variation of the other parameters of the model (except for $n_s$, which appears to remain constant). We then write
\begin{equation}
\label{eq:new_global_mass_mdl}
\begin{split}
	\alpha =& \alpha_a + \alpha_b \ (1 - N_{\%} ) \; , \\
	 \log_{10} f_1 =& \log_{10} f_{1a} + \log_{10} f_{1b}  \  (1 - N_{\%} ) \; , \\
	 \log_{10} f_{\mathrm{knee}} =&  \log_{10} f_{\mathrm{knee},a} +  \log_{10} f_{\mathrm{knee},b}  \  (1 - N_{\%} ) \; ,  \\
	 \log_{10} f_2 =&  \log_{10} f_{2a} +  \log_{10} f_{2b}  \  (1 - N_{\%} )  \; ,
\end{split}	 
\end{equation}
which relates all parameters of the PSD model of~\cref{eq:galfit_strain} to a linear combination of two extra parameters and the $(1 - N_{\%})$ term. This will allow us to estimate the set of new parameters, which would be universal for any population with a given $m_{\mathrm{gal}}$, since $m_{\mathrm{gal}}$ can be related directly through the $(1 - N_{\%})$ term.  

\begin{figure}
  \begin{center}
	\includegraphics[width=\linewidth]{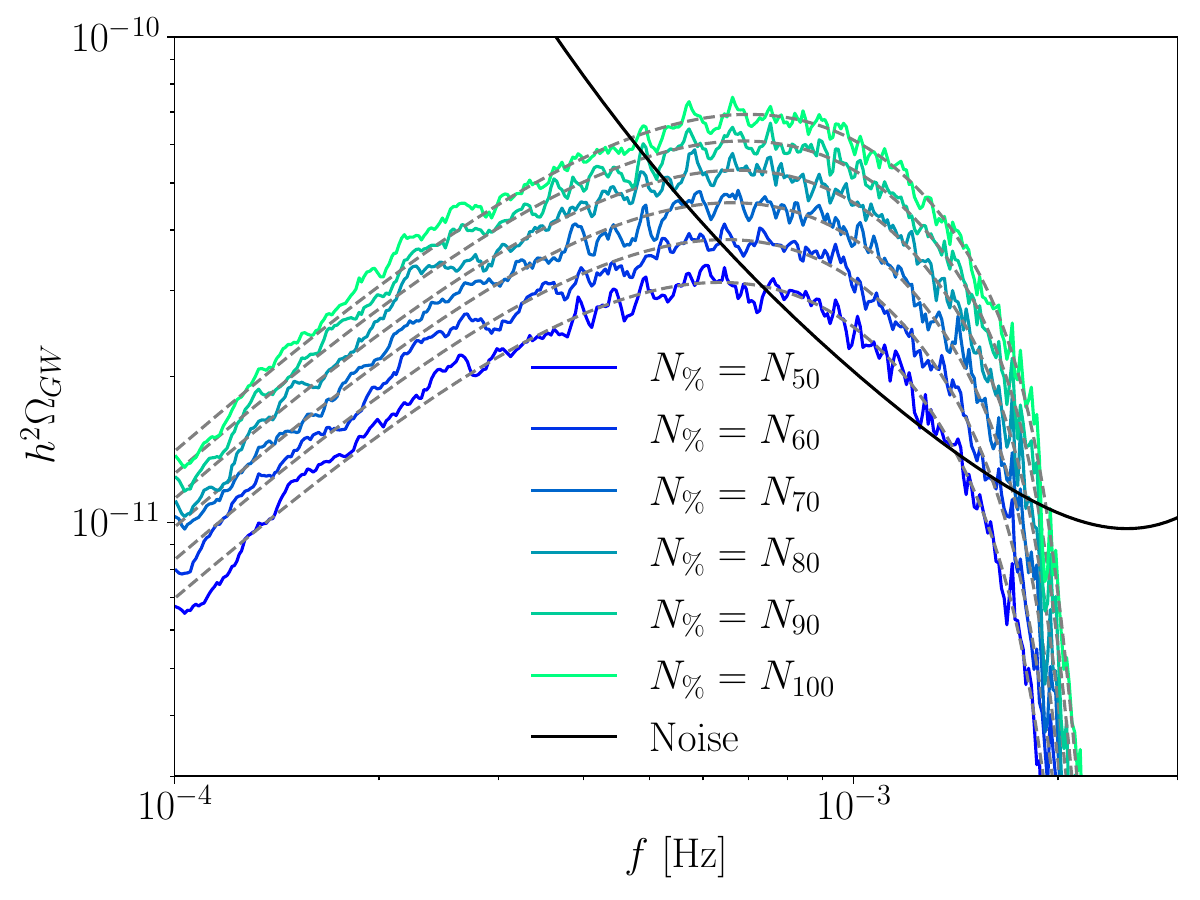}
\end{center}
	\caption{\label{fig:Global_fit} The result of the joint fit of the six datasets with the different number of sources $N_\%$ (colored solid lines). The gray dashed line represents the model of~\cref{eq:galfit_strain}, evaluated at best fit parameters, using their relations of~\cref{eq:new_global_mass_mdl}. The black solid curve represents \textit{LISA}'s instrumental noise.}
\end{figure}
\begin{figure}
\begin{center}
	\includegraphics[width=.65\linewidth]{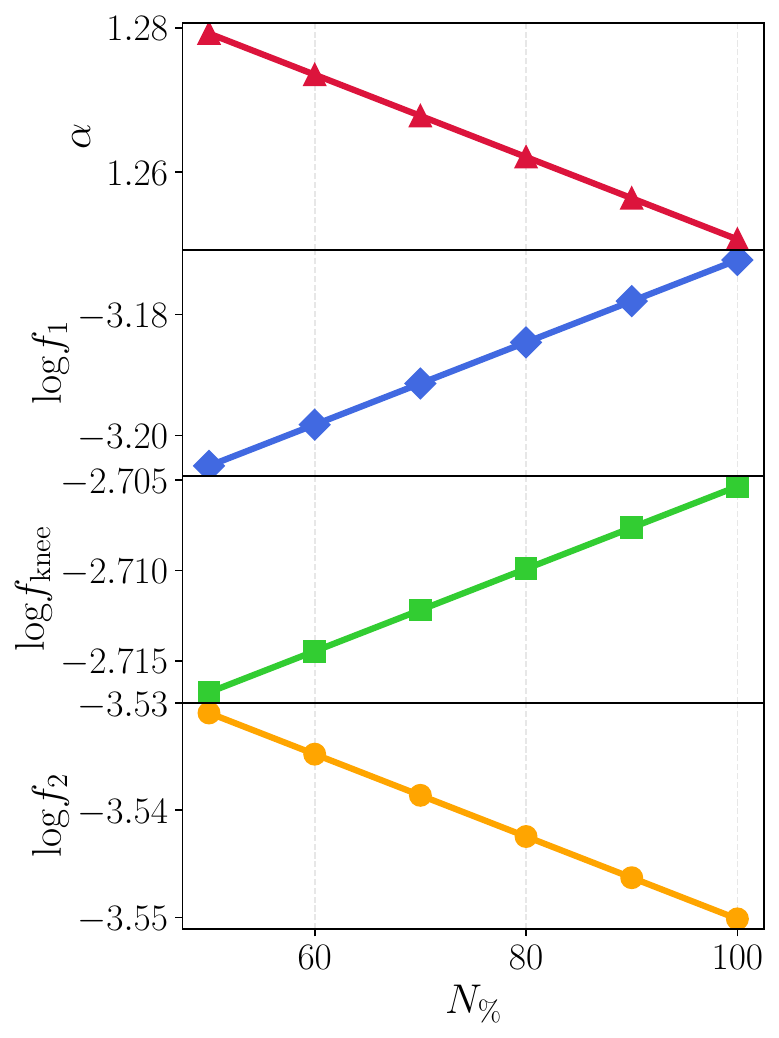}
\end{center}
	\caption{\label{fig:mass_params} The dependence of the parameters of the spectral model of~\cref{eq:general_likelihood}, on the number of sources $N_\%$ contained in the dataset. See text for details. }
\end{figure}

\begin{figure*}
    \makebox[\textwidth][c]{ \includegraphics[width=1.2\linewidth]{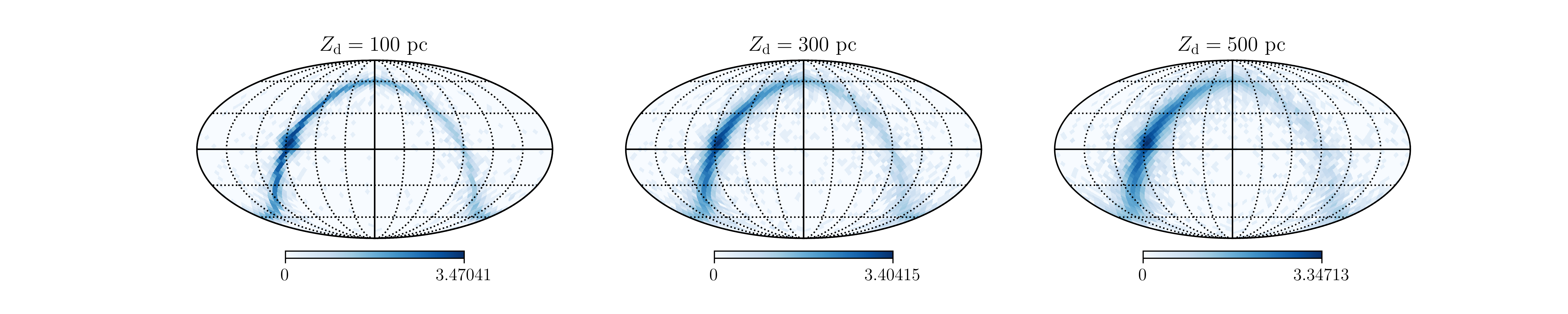} }
 	\caption{ \label{fig:mollview} The density of resolved sources, for the three cases of simulated data investigated in~\cref{sec:Shape_Impact}. We have simulated the same population for different height scale $Z_\mathrm{d}$ and computed the corresponding resolvable sources following the methodology of~\cref{sec:foreground_estimation}. The color bar units are in $\log_{10}(1+N_\mathrm{sources})$, where $N_\mathrm{sources}$ is the number of sources for each case (see~\cref{tab:zsources}).} 
\end{figure*}

Having defined the relations of~\cref{eq:new_global_mass_mdl}, we can now perform a global fit for the six datasets using the model of~\cref{eq:galfit_strain} for each, evaluated at the parameters estimated from~\cref{eq:new_global_mass_mdl}. The next step is to define a likelihood function, which has the form of
\begin{equation}
    \label{eq:mass_likelihood}
	-2 \ln \mathcal{L} = \sum_{k,ij} w^k_{ij} \left(1 -  \frac{D^k_{ij} } {D_{ij}^{th}(f^k_{ij}, \vec{\theta}) }  \right)^2  \; ,
\end{equation}
where $D^k_{ij}, f^k_{ij} $ and $w^k_{ij}$ are, respectively, the data (where the noise is included at face value), the frequencies and the weights as defined in~\cref{sec:analytic_model}. On the other hand, $D_{ij}^{th}(f^k_{ij}, \vec{\theta}) $ denotes the model for the data (which depends only on the signal parameters $\vec{\theta})$, \emph{i.e.} the noise is again included at face value, without any explicit parameter dependence. We include the noise, even if non-parametric, in this estimate in order to give less weight to frequency ranges where the amplitude is very low (in particular, the large frequency ones, which have larger weights and would dominate the estimate). In a more realistic setting, the noise should also depend on a set of parameters, which, in a fully consistent Bayesian framework, should be measured simultaneously with the signal parameters. 

\subsubsection{Numerical results}

A plot of the obtained datasets corresponding to $N_{\%} = \{ 0.5, 0.6, 0.7, 0.8, 0.9, 1\}$ (or equivalently to the percentage of the total stellar mass $m_{\rm G} = 8.2\times10^{10}\,$M$_\odot$) is shown in~\cref{fig:mass_dep}. From the right panel of the plot of this figure, it is clear that after re-scaling the amplitude of the stochastic signals with $1/N_{\%}$, there is significant overlap between the simulated data. This immediately points to an underlying relation of the total number of binaries and the overall level of the signal PSD. 

The fit of the spectral shape, as described in the previous~\cref{sec:fitting_the_shape}, is numerically tackled by using a Metropolis-Hastings MCMC sampler (see the Data Availability section for details), which maps the parameter space of the likelihood function in~\cref{eq:mass_likelihood}.
The results of the analysis, are shown in~\cref{fig:Global_fit} and~\cref{fig:mass_params}. In~\cref{fig:Global_fit}, we present the different residual foreground confusion signals (colored solid lines) for the given $N_{\%}$ catalogues, and the best-fit spectral model of~\cref{eq:galfit_strain} plotted on top of the data (dashed gray lines). The result of the fit yields the set of parameters that enables us to relate each dataset through their corresponding catalogue. This is shown in~\cref{fig:mass_params}, where the evolution of the parameters~\footnote{Notice that the full set of parameters $\vec{\theta}$ used for this analysis also includes a global amplitude parameter $\log_{10}(h^2 \Omega_{\rm gw}^*)$ and a global low-frequency tilt parameter $n_s$.} $\vec{\theta}_{\rm P} \equiv \{\alpha,\, \log f_1,\, \log f_\mathrm{knee},\, \log f_2 \}$ are given with respect to  $N_{\%}$. Interestingly, both $\log f_1$ and $\log f_{\rm knee}$ increase $N_{\%}$, meaning that the high-frequency edge of the spectrum moves to higher frequencies. This is expected as, by increasing the number of sources, the overall confusion noise amplitude increases, making it more difficult to subtract individual events. Similarly, the decrease in $\alpha$ shows that for larger $N_{\%}$, having more sources (and a smaller relative number of subtracted sources) we expect a milder loss in stochasticity. Finally, the decrease in $\log f_2$ shows that, independently on the previous effects, as a certain frequency is reached the identification of individual sources is so efficient that a sharp cut-off stays present rather independently on $N_{\%}$.

In order to get an estimate of the stellar mass need to generate the `observed' spectral shape of the confusion foreground, using an alternative approach to previous studies, our result can be used as a look-up-table for a chosen catalogue with a set binary evolution model and star formation history of the Galaxy (cf. sec.~\ref{sec:background}). Naturally, this result depends on the adopted binary population model, but the methodology presented here, in principle, could be applied to any given Galactic population that generates this confusion signal for \textit{LISA}. It is also important to notice that the spectral model adopted in~\cref{sec:analytic_model} is flexible enough to account for different population models. For example, it could be adapted for cases, where the spectral shape cannot be captured entirely by~\cref{eq:galfit}, such as the observationally-driven model explored in~\cite{KorolHallakoun2021}.

As a final step, after the parametrization in~\cref{eq:new_global_mass_mdl} is set and the best fit parameters are determined by using MCMC methods, we reverse the argument to constrain the $N_{\%}$ parameter, or equivalently, the uncertainty on the Galactic mass. For this purpose, we begin with the assumption that the spectral model of the confusion signal is known. Then, the parameters we had previously estimated are considered known as well, and can be picked depending on their relation to $N_{\%}$ directly from~\cref{fig:mass_params}. We perform a Fisher forecast on the simulated datasets, assuming only the amplitude $\log_{10}(h^2 \Omega_{\rm gw}^*)$ and $N_{\%}$ parameters. We choose this set of  parameters for the sake of simplicity, and due to the evident dependence of the number of total binaries in the population to the overall spectral amplitude (cf.~\cref{fig:mass_dep}). For each considered value of $N_{\%}$, we retrieve percentage-level relative errors (see~\cref{tab:npercent}). As expected, the uncertainty is shrinking for datasets with larger $N_{\%}$, corresponding to a stronger stochastic signal. With this information at hand, and since the number of binaries scale proportionally to the overall stellar mass, the same quantities should apply to $m_{\rm G}$. We report this result in~\cref{tab:npercent}. 
\begin{table}
  \begin{center}
    \caption{The recovered uncertainty for the $N_{\%}$ parameter (and therefore the $m_{\rm G}$), after computing a Fisher Matrix of the two parameters of interest $\log_{10}(h^2 \Omega_{\rm gw}^*)$ and $N_{\%}$ (see text for details). The resulting relative uncertainty is at the percentage level, and as expected, it is decreasing for stronger stochastic foreground signals (or larger $N_{\%}$).}
    \label{tab:npercent}
    \begin{tabular}{ c|c|c } 
      \textbf{$N_{\%}$} & \textbf{$m_{\rm G}~[\times10^{10} \mathrm{M}_\odot]$} &  Relative error $1\sigma$\\
      \hline
      $N_{50}$ & $4.1$ & 0.017  \\
      $N_{60}$ & $4.92$ & 0.013  \\
      $N_{70}$ & $5.74$ & 0.010  \\
      $N_{80}$ & $6.56$ & 0.008  \\
      $N_{90}$ & $7.38$ & 0.007   \\
      $N_{100}$ & $8.2$& 0.006  \\\hline
    \end{tabular}
  \end{center}
\end{table}

%
%
\subsection{Probing the shape of the Galactic disc}
\label{sec:Shape_Impact}

\begin{figure*}
 \begin{center}
    \includegraphics[width=.47\linewidth]{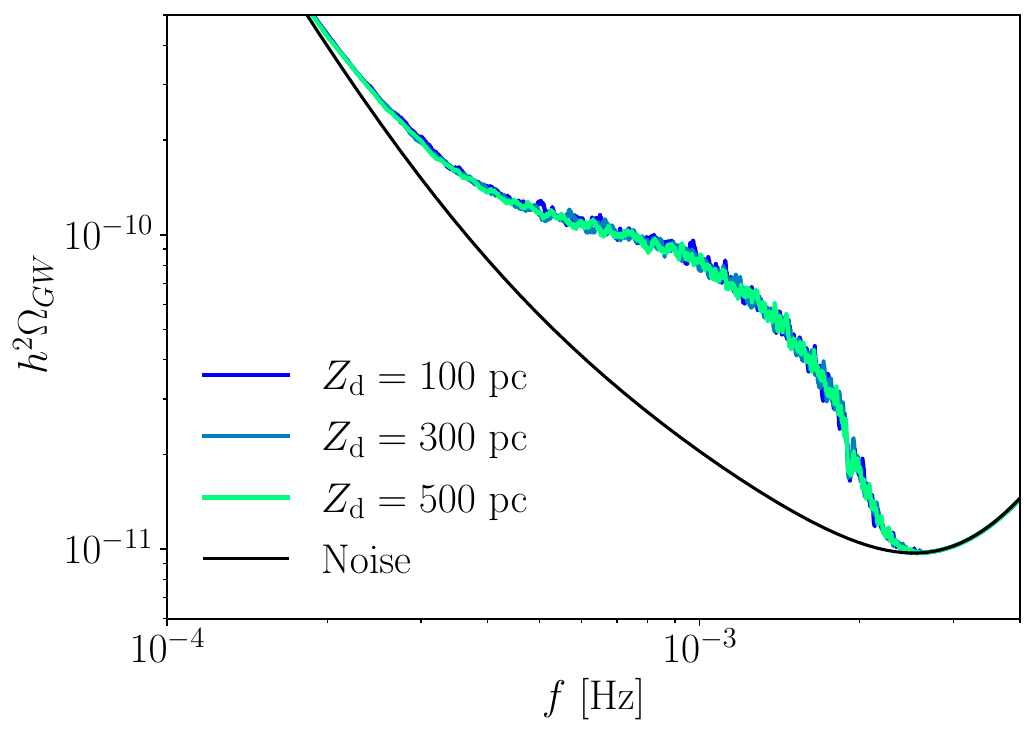}\hspace{1.cm}
    \includegraphics[width=.34\linewidth]{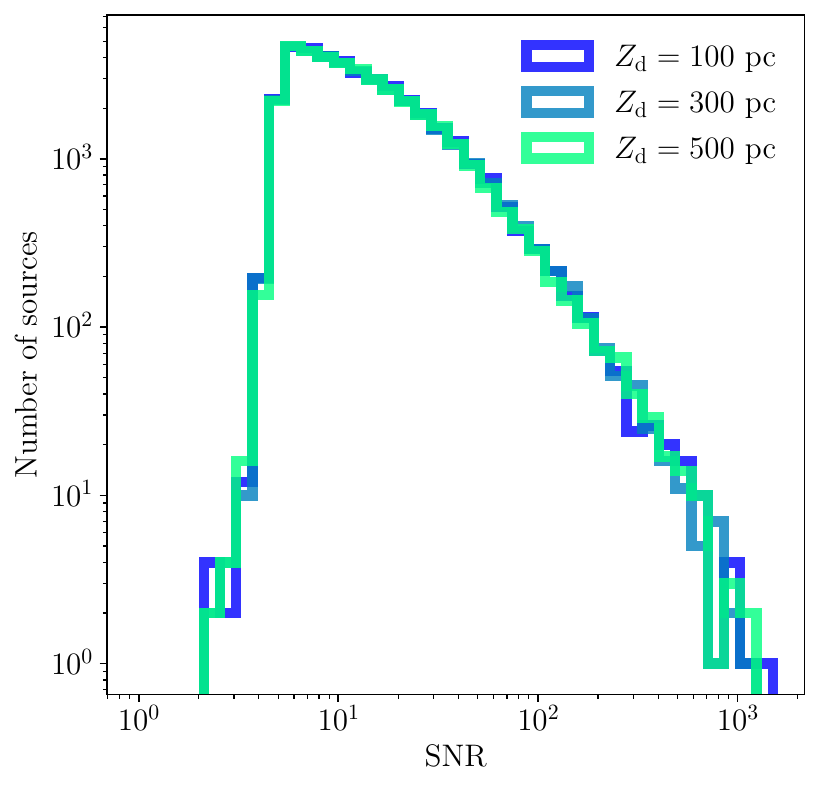}
\end{center}
    \caption{{\it Left}: Unresolved foreground signals for the A channel with instrument noise for the 3 datasets. The \textit{LISA} instrument noise PSD is represented with the black line~\citep{SciRD}. As expected, the foreground signal is very similar for all cases considered (see text for details). {\it Right}: Histogram of the number of subtracted sources for each dataset, assuming that sources with SNR above $5$ are resolvable. The total number of the resolvable sources is also mentioned in \cref{tab:zsources}. }
    \label{fig:Z_5}
\end{figure*}

We now focus on the impact of different Galactic disc shapes to the overall recovered foreground signal, as well as the resolved number of sources for each case. We restrict our analysis to the scale height of the disc $Z_\mathrm{d}$, which has a greater impact on the spatial distribution of DWDs, compared to the disc scale length $R_{\rm d}$ (cf. equation~\ref{eqn:disc}). Specifically, we use three different simulated catalogues, keeping the intrinsic parameters of the sources constant, while only varying the  disc scale heights using $Z_\mathrm{d} = \{100,\, 300,\, 500\}~\mathrm{pc}$, respectively. Our motivation is as follows. Different stellar populations are characterized by different scale heights: main sequence stars are born in the inner part of the Galactic disc (within the thin discs), whereas older stellar populations exhibit larger scale heights, proportional to the vertical velocity dispersion \citep[e.g.][]{mac17}. As DWDs dominate the \textit{LISA} band at lower frequencies, it is expected that the shape of the Galactic confusion foreground is linked to their density and spatial distribution. 

Following the methodology described in~\cref{sec:foreground_estimation}, we estimated the resolvable sources and residual foreground signal for each value of $Z_\mathrm{d}$. As already mentioned, for each of the catalogues, the number of sources was kept the same, and only the spatial distribution has been modified.  A sky density plot of the resolvable sources density for each case is shown in~\cref{fig:mollview}, where the effect of the scale height is quite evident. The equivalent PSDs of the foreground signals are presented in the left panel of~\cref{fig:Z_5}, where different data sets are practically indistinguishable by eye. The latter result is in line with earlier studies. For example, ~\cite{Ruiter2009} also found that subtraction of `loud' and resolvable sources affects very weakly the low-frequency, confusion-limited signal. One would expect that with increasing scale height, the resulting average distance to the binaries would also increase; thus, the capabilities of \textit{LISA} to detect those sources with high confidence would also be affected. With our analysis here, we show that the number of resolvable sources does not change significantly for the three considered cases of $Z_\mathrm{d}$.
\begin{table}
  \begin{center}
    \caption{The recovered sources for the three catalogues with different scale heights. For all three cases we notice very similar number of resolved sources, with similar extraction SNR values (see right panel of~\cref{fig:Z_5}).}
    \label{tab:zsources}
    \begin{tabular}{l|c} 
      \textbf{Catalogue} & \textbf{Number of recovered sources} \\
      \hline
      $R_z=100~\mathrm{pc}$ & 38342 \\
      $R_z=300~\mathrm{pc}$ & 38498 \\
      $R_z=500~\mathrm{pc}$ & 38848 \\\hline
    \end{tabular}
  \end{center}
\end{table}
This is made clear in the right panel of~\cref{fig:Z_5}, where we plot the histograms of the subtracted sources as a function of their recovered SNR. For all cases, we also classify as detected almost the same number of sources, which is shown in~\cref{tab:zsources}. A small variation of the recovered sources could be explained by the decrease in the density of sources (as $Z_\mathrm{d}$ increases), which in turn would boost the detectability of a small percentage of them. However, more detailed simulations are needed in order to verify this effect. 

Another way of cross-validation  is to compare the predicted error bars of the recovered sources, as estimated via the Fisher Information Matrix analysis. As described in~\cref{sec:foreground_estimation}, for each of the catalogues of recovered sources corresponding to a given $Z_\mathrm{d}$, we can estimate the predicted covariance matrix of the parameters as the lower Cramer-Rao bound~\citep{PhysRevD.77.042001}. The result is shown in \cref{fig:zfisher}, where we can see that, in general, the estimated errors on the parameters are not heavily dependent on the actual $Z_\mathrm{d}$ of the simulated catalogue (at least for the values of $Z_\mathrm{d}$ considered in our analysis). 

It is worth commenting here, that for this investigation we have again assumed the nominal mission duration of $\mathrm{T}_\mathrm{obs} = 4~\mathrm{years}$. In the extreme case of much longer duration ({\it i.e. } $\mathrm{T}_\mathrm{obs} \sim 10~\mathrm{years}$), and depending on the given population model, the resulting stochastic signal would be much lower in amplitude due to the increased number of resolvable sources~\citep{Karnesis2021tsh}. Therefore, the impact of scale height to the overall signal would be even weaker, and thus more challenging to detect. More reliable information could be retrieved from the resolved sources instead. With $\mathrm{T}_\mathrm{obs} \sim 10~\mathrm{years}$, and with good localization for a great fraction of them, the scale height could be more straightforward to determine.

%
%
%
%
\section{Population-based Inference using the resolved sources}
\label{sec:hierarchical}

As we have seen in previous sections, and also shown from previous works, \textit{LISA} is going to detect and resolve (i.e. distinguish individual signals) a small percentage  of ultra compact binaries in the Galaxy (of order $\sim 0.1$\,per cent). Even so, the actual number of resolved sources will amount to $\sim10^4$, as predicted by a number of different population models \citep[for a review see][]{amaro22}. These are challenging to detect with electromagnetic observations, thus the \textit{LISA}'s sample will provide a unique opportunity to probe the properties of the underlying population model at shortest orbital periods. To do that, one can employ a hierarchical Bayesian approach, as proposed in~\citep[e.g.][]{adams2012, LIGOScientific:2020kqk, PhysRevD.104.082003, PhysRevD.98.083017, PhysRevD.98.063001}.

In this framework, we begin by writing 
\begin{equation}
    \label{eq:bayes}
	p(\vec{\theta} | d ) = \frac{\mathcal{L}(d | \vec{\theta}) p(\vec{\theta} )}{p(d)},
\end{equation}
where $\mathcal{L}(d | \vec{\theta})$ the likelihood of the data given the parameters $\vec{\theta}$, and $p(\vec{\theta} )$ the prior belief we might have on those parameters. The $p(d) \equiv \int_\Theta  \mathcal{L}(d | \vec{\theta}) p(\vec{\theta} ) \mathrm{d}\vec{\theta}$ is the marginal likelihood, or {\it evidence}, which acts as a normalization constant in most parameter estimation problems. The evidence of a model, is essentially providing us with the capabilities of the particular model on the data. Thus, it can be used for model selection problems. In general, higher-dimensional models are penalized due to the increased volume in the parameter space (embodied in the priors $p(\vec{\theta} )$ of the parameters)~\citep{gelmanbda04}. Assuming Gaussian properties for the noise, we can write the logarithm of the likelihood as:
\begin{equation}
    \label{eq:general_likelihood}
	\ln \mathcal{L}(d | \vec{\theta}) \propto -\frac{1}{2}\left( d - h(\vec{\theta}) \Big| d - h(\vec{\theta}) \right).
\end{equation}
Here, $d$ is the given dataset and $h(\vec{\theta})$ is the given model that depends on a set of parameters $\vec{\theta}$. In fact, here $\vec{\theta}$ will represent all parameters of the recovered binaries. We then find a solution for the maximum likelihood via setting $\partial \mathcal{L} / \partial \theta_i = 0$, and make use of the FIM, $F_{ij}$, which will yield 
\begin{equation}
    \label{eq:fim_likelihood}
	\ln \mathcal{L}(d |  \vec{\theta}) \propto -\frac{1}{2}(\theta_i - \theta_i^\mathrm{ML} ) (\theta_j - \theta_j^\mathrm{ML}) F_{ij}, 
\end{equation}
where $\vec{\theta}^\mathrm{ML}$ denotes the Maximum Likelihood parameters. This last approximation is very useful, because it allows us to use the  Gaussian statistics assumptions of the FIM instead of computing directly \cref{eq:general_likelihood} for all $\mathcal{O}(10^4)$ binaries that we recover with each catalogue. We are therefore able to ease the computational burden, by drawing the $\Delta\vec{\theta} = \vec{\theta} - \vec{\theta}^\mathrm{ML}$ from a multivariate Gaussian with a covariance matrix $\Sigma_{ij} = F_{ij}^{-1}$. This approach is also very convenient because the FIMs for all the recovered sources are pre-computed when the procedure described in~\cref{sec:foreground_estimation} is performed. 

\subsection{Astrophysical priors}

Here we chose to investigate some parameters, particularly interesting from the astrophysical prospective. These are the frequency $f_0$, chirp mass ${\cal M}$ and time to coalescence $\tau_c$ distributions. As stated above, \textit{LISA} will provide a complete sample of DWDs up to frequencies of $\sim 2\,$mHz (or equivalently the orbital periods of up to $\sim 30\,$min). DWDs with such  short orbital periods are technically challenging to detect with electromagnetic telescopes, due to the compact size of these binaries and white dwarf stars, and their unique spectral characteristics. As a consequence, the available sample is highly biased and incomplete \citep[e.g.][]{Kupfer2018,reb19}. Similarly, it is particularly hard to constrain the DWD chirp mass distribution with currently available observations. This is due to the many biases in the known DWD sample, and to the fact that only lower limit estimates are available for the masses of most of the photometric secondaries, which often remain unseen \citep[see figure~1 of][]{KorolHallakoun2021}. Binaries' GW frequencies and chirp masses measured by \textit{LISA} can be combined to obtain the distribution of their coalescence time, the integral of which represents a measure of the merger rate for these binaries. It is important to mention that the distributions of $f_0$, ${\cal M}$ and $\tau_c$ will be of fundamental importance for our understanding of the supernovae type Ia progenitor problem \citep[e.g.][]{Ruitr2009snIa,Maoz2012,Toonen12,Shen2012}. For the sake of convenience ({\it i.e.} more straightforward model for the chirp mass prior), for this application we use the population of~\cite{KorolHallakoun2021}.

%
%
\subsubsection{The chirp mass prior}

We begin with the estimates of the FIM calculated in the procedure described in \cref{sec:foreground_estimation}. From the waveform parameters, we can construct the relative errors on the chirp mass ${\cal M}$ of each object using linear error propagation as: 
\begin{equation}
    \label{eq:mc_sigma}
	\frac{\sigma_{\cal M}}{{\cal M}} = \sqrt{ \left( \frac{11}{5} \frac{\sigma_f}{f} \right)^2 + \left( \frac{3}{5} \frac{\sigma_{\dot{f}}}{\dot{f}} \right)^2 + \frac{33}{25} \left(\frac{\sigma_f}{f} \right) \left( \frac{\sigma_{\dot{f}}}{\dot{f}} \right) \rho_{f\dot{f}}},
\end{equation}
where a dot denotes derivative with respect to time, and $\rho_{f\dot{f}}$ is the correlation coefficient between the $f$ and $\dot{f}$ parameters. The second step is to adopt a model for the chirp mass distribution. While, as expected, there are selection effects that need to be taken into account, for each different catalogue case we usually resolve enough sources that allow us to constrain at least the lower-mass part of the mass spectrum (${\cal M} \le 0.7$). In our work here, we focus on that region of chirp mass prior, which is more straightforward to characterize~\cite{adams2012}. Motivated by the shape of the chirp mass distribution of our simulated population, for its probability function we adopt a model that follows a set of broken power-laws in logarithmic space as
\begin{equation}
\label{eq:m_pdf}
\begin{split}
	p({\cal M}) =&  10 ^{a_1} \left(\frac{{\cal M}} {p_1} \right)^{n_1} {\mathcal H}\left({\cal M}_1 - {\cal M}\right)  \\ 
	            &  + 10^{a_2} \left(\frac{{\cal M}} {p_2}\right)^{n_2} {\mathcal H}\left({\cal M} - {\cal M}_1\right) {\mathcal H}\left({\cal M}_2 - {\cal M}\right)  \\ 
	            &  + 10^{a_3} \left(\frac{{\cal M}}{p_3}\right)^{n_3} {\mathcal H}\left({\cal M} - {\cal M}_2\right), 
\end{split}
\end{equation}
with ${\mathcal H}$ being the Heaviside step function, ${\cal M}_1$ and ${\cal M}_2$ are the conjunction points between the three different parts of the power laws, $n_i$ are the different slopes, and $p_i$ are constants. Since we pose that the three-piece power laws is a continuous function, the $a_i$ are related as
\begin{eqnarray}
	a_1 &=&  a_2 + \log\left[ \left(\frac{{\cal M}_1}{p_1}\right)^{-n_1} \left(\frac{{\cal M}_1}{p_2}\right)^{n_2} \right] \\
	a_3 &=&  a_2 + \log\left[ \left(\frac{{\cal M}_2}{p_3}\right)^{-n_3} \left(\frac{{\cal M}_2}{p_2}\right)^{n_2} \right].
\end{eqnarray}
Additionally, since we work with probability densities, the $a_i$ parameters are not really relevant for this analysis. As we will see in the sections below, only the slopes $\{n_1,\, n_2,\, n_3\}$ are going to be useful for the characterization of the chirp mass Probability Density Function (PDF). 

%
%
\subsubsection{The frequency and time of coalescence priors}
\label{sec:f0_and_tc} 

Assuming that in the \textit{LISA} frequency band the evolution of the binaries is only dependent on the emission of GWs, we expect the $f_0$ to follow a power-law, with a slope of $n=2/3$. This can also be verified by the leftmost panel of figure~\ref{fig:pdfs}, which shows which part of the complete distribution can be accessed by the population of resolvable sources for this particular simulated catalogue. The model simply reads as
\begin{equation}
    \label{eq:f0}
	p(f_0) = 10^\alpha \left( \frac{f_0}{f^\ast}\right)^{-n},
\end{equation}
with $10^\alpha$ being the amplitude, $n$ the slope, and $f^\ast$ the pivot frequency set to $f^\ast=10^{-5}$. In fact, just as in the case for the chirp mass, since we are working with densities, the amplitude parameters are normalized out. 

The time of coalescence $\tau_c$, and its measurement error, is a parameter that can be derived from the rest of the waveform parameters. In order to do that, we need to make a couple of simplifying assumptions. First, we begin by assuming that the binaries have equal masses. This assumption is motivated by the fact the mass ratio of our synthetic DWD peaks at 1 \citep[e.g. as visible in figure~8 of][]{Korol2017}. We can then derive the total mass $m_\mathrm{tot}$ of each binary from the chirp mass as
\begin{equation}
    \label{eq:mtot}
	m_\mathrm{tot} = \frac{{\mathcal M}}{\left(1/4\right)^{3/5}}.
\end{equation}
Then, for circular binaries we can approximate the value of $\tau_c$ by numerically integrating~\citep[e.g. p. 170 of][]{Maggiore}
\begin{equation}
    \label{eq:dfdt}
	\frac{\mathrm{d}f_\mathrm{gw}}{\mathrm{d}t} = \frac{96}{5}  \pi^{8/3} \left( G {\mathcal M} / c^3 \right)^{5/3} f_0 ^{11/3}.
\end{equation}
Here, the $f_0$ frequency is the given source emission frequency as measured by \textit{LISA}. The integral is computed up until the emission frequency at the Innermost Stable Circular Orbit  $f_\mathrm{isco}$, which is given by 
\begin{equation}
    \label{eq:isco}
	f_\mathrm{isco} = \frac{c^3}{12\pi\sqrt{6}Gm_\mathrm{tot}}.
\end{equation}
However, for white dwarfs $f_{\rm isco}$ can be set at $\sim 30\,$mHz as they start interacting when one of the stars overfills its Roche lobe, which may lead directly to the merger \citep[][]{Shen2015}. 

By looking at the given population properties (see \cref{fig:pdfs}), the $\tau_c$ parameter distribution could also be characterized by power-law function, in the same manner as we did for the $f_0$ parameter. Then we write
\begin{equation}
    \label{eq:tc}
	p\left(\tau_c\right) = 10^\gamma \left( \frac{\tau_c}{\tau^\ast}\right)^{-n_{\tau_c}},
\end{equation}
where $10^\gamma$ is the amplitude, $n_{\tau_c}$ the slope, and $\tau^\ast$ the pivot point of the power-law, which is just a convention. Here we fix $\tau^\ast = 10^{24}$. Again, the $\gamma$ parameter is not relevant in the analysis, because we are fitting densities. The predicted error bars on $\tau_c$ can be derived by computing the error propagation rules from the measured $\{ {\mathcal M}, f_0, \dot{f}_0 \}$. Since $\tau_c$ is a derived parameter, we could simply estimate the $n_{\tau_c}$ parameter independently from the rest, but one could just include it in the global analysis for a joint fit. The joint analysis we performed is described in the following section.


%
%
\subsection{Joint fit under a hierarchical Bayesian model}
\label{sec:joint_analysis}

So far, we have defined the PDF functions of three parameters of interest, the chirp mass ${\mathcal M}$, the emission frequency $f_0$, the time of coalescence $\tau_c$, and their corresponding hyperparameters. From eqs.~(\ref{eq:tc},~(\ref{eq:f0}), and~(\ref{eq:m_pdf}), those are the $\vec{\beta} = \{ n_{\tau_c},\, n_1,\, n_2,\, n_3,\, n_{f_0} \}$. With this information at hand, and beginning from what we had written in~\cref{eq:bayes}, we can write a hierarchical Bayesian model as
\begin{equation}
    \label{eq:hierarchical}
	p(\vec{\theta}, \vec{\beta} | d ) = \frac{\mathcal{L}(d | \vec{\theta}) p(\vec{\theta} | \vec{\beta} ) p(\vec{\beta})}{p(d)},
\end{equation}
where we have elevated $\vec{\beta}$ as hyperparameters, and also introduced a prior distribution $p(\vec{\beta})$ for them as well. Or in other words, we have defined the $p(\vec{\theta} | \vec{\beta} )$ term that describes the waveform parameters prior, for distribution models described by hyperparameters $\vec{\beta}$. Now, the evidence $p( d )$ is given by marginalizing over the full parameter space of $\vec{\theta}$ and $\vec{\beta}$. It is useful to remind here, that for the computation of the likelihood that appears in the above~\cref{eq:hierarchical}, we adopt the approximation of~\cref{eq:fim_likelihood} where we utilize the already computed FIMs for the resolved sources, with $\Sigma = F_{ij}^{-1}$ for the waveform parameters. 
\begin{figure*}
 	\includegraphics[width=1.\linewidth]{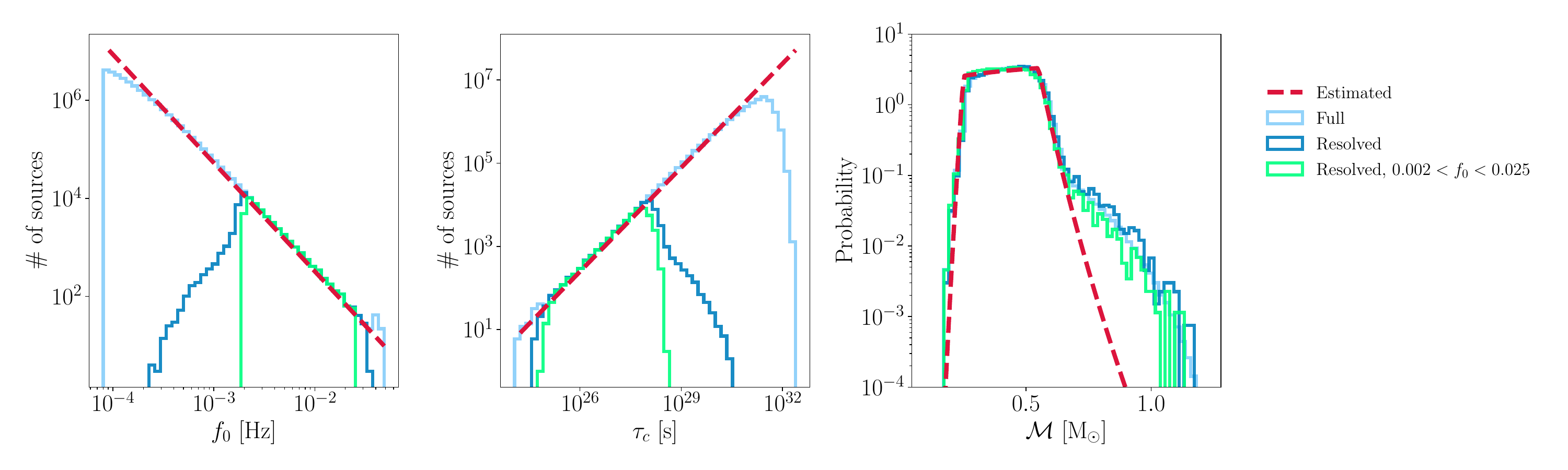}
 	\caption{ \label{fig:pdfs} The histograms of the population parameters, compared with the histograms of the sub-set catalogue of resolvable sources. With light blue we show the full catalogue histogram, with dark blue the recovered subset, while with the dashed red line we show the best-fit model. {\it Left}: The PDF of the main emission frequency $f_0$. The fit is performed on the regions where the population is well recovered (green data), but extended over the complete parameter space (see text for details). {\it Middle}: The PDF for the logarithm of the time of coalescence. {\it Right}: The chirp mass PDF case. The distribution of the recovered sources with characteristic frequency above $0.3~\mathrm{mHz}$, follows closely the true one, originating from the complete population (green curve). The fit is satisfactory for ${\mathcal M}\lessapprox 0.7$ due to the sufficient number of resolved sources with measurable chirp mass (see discussion in section~\ref{sec:joint_analysis} for details).}
 \end{figure*}

The second step would be to impose a set of priors on the hyperparameters, which of course will depend on the given population model. Starting with the chirp mass ${\mathcal M}$ case, we first have to account for possible selection effects of the catalogue of the resolvable sources, which are quite evident from the right panel of~\cref{fig:pdfs}. In this figure, we compare the ${\mathcal M}$ PDFs of both the full population (in light blue color), and the sub-set catalogue that contains the resolvable sources (in dark blue color). Both functions agree well for the lower part of the mass spectrum, but as expected, they diverge for ${\mathcal M}\lessapprox 0.7~M_\odot$. The reason is that the confusion stochastic ensemble signal dominates the frequency ranges between $0.1$ to $3$~$\mathrm{mHz}$, and therefore causes fewer detectable sources, and with larger estimated error-bars. However, for this specific population (that was described in~\citep{KorolHallakoun2021}), after performing the FIM analysis mentioned in section~\ref{sec:foreground_estimation}, we recover a considerable percentage of sources with measurable chirp mass. In particular, the $15$\,per cent of the resolved sources, will have a chirp mass relative error lower than $10$\,per cent. These provide us with enough weight through \cref{eq:mc_sigma} in order to be able to recover sufficient information about the chirp mass distribution\footnote{As a cross-validation test, we have explored the limitations of this methodology by attempting to estimate the hyperparameters of the chirp mass distribution for different number of resolvable sources. In particular, we have repeated the analysis by considering sources with different levels of mass measurement errors. We have found that the chirp mass prior is sufficiently reconstructed for ${\mathcal M}\gtrapprox 0.7$ by utilizing sources up to 100\% of their relative errors. However, this figure is naturally dependent on the given binary population model under study.}.

Considering the above, the features of the PDF of ${\mathcal M}$ above $0.7~\mathrm{M}_\odot$ are not accessible from this analysis alone. For the above reasons, we restrict the fit for $0 \leq {\cal M} \leq 0.75$, and we adopt uniform priors for the parameters of~\cref{eq:m_pdf} as $n_1 \sim {\mathcal U}[-10,\, 50]$, $n_2 \sim {\mathcal U}[-50,\, 10]$, and $n_3 \sim {\mathcal U}[-100,\, -1]$. We also choose to keep ${\cal M}_1=0.25\,$M$_\odot$ and ${\cal M}_2=0.75\,$M$_\odot$ constant for the sake of simplicity. We recall that ${\cal M}_1$ and ${\cal M}_2$ represent constants of our astrophysically-motivated prior (see \cref{eq:m_pdf}).

The $f_0$ and $\tau_c$ cases are more straightforward. As already mentioned in~\cref{sec:f0_and_tc}, we have adopted power-law models. We have chosen to use a broad uniform prior of the slope parameters $n_{f_0} \sim {\mathcal U}[-5,\, 5]$ and $n_{\tau_c} \sim {\mathcal U}[-10,\, 10]$. In order to avoid selection bias effects, we choose to fit the hyperparameters at regions where the hyperprior distributions are well characterized by our adopted models. For that reason, we restrict the fit for sources with main emission frequency between $2 < f_0 < 25~\mathrm{mHz}$. Finally, as in~\cref{sec:Mass_impact}, we use MCMC methods in order to map the posterior surface of the hyper parameters.  

The result is shown in \cref{fig:pdfs}, where the hyper-distribution models are plotted at best-fit hyper-parameters (dashed red lines). Together, we show the histograms of the complete population for comparison. For $\tau_c$ and $f_0$ the distributions recovered are accurately describing the population of the resolved sources (dark blue). There is a noticeable artifact on the left panel of~\cref{fig:pdfs}, which represents the $f_0$ distribution. The histogram that corresponds to the recovered sources subset is vanishing for $f_0 \gtrapprox 0.02~\mathrm{Hz}$. This is caused by the data generation procedure, where we need to select a given cadence for the simulation. In our case, in order to save computational resources, we have chosen $\Delta t = 15~\mathrm{s}$, which brings the Nyquist frequency down to $0.033~\mathrm{Hz}$. This means that sources with higher characteristic frequencies were inaccessible in the analysis. As in the case of $f_0$ and $\tau_c$, the chirp mass function is not fully recovered, due to selection effects that were discussed above in this section. The corner plot of the posteriors of the hype-parameters is shown in~\cref{fig:posteriors}. As expected, no correlations where found between $n_{\tau_c}$ and the rest of the parameters, since $\tau_c$ is a derived quantity.

Considering the above results, we can conclude that given the assumptions about the mission (overall measurement duration and noise PSD levels), we can indeed probe some of the astrophysically interesting properties of the particular compact Galactic DWD population model. In our analysis here we focused only on the chirp mass, the main frequency and time of coalescence functions, but one could extend it in the overall shape of the Galaxy, as demonstrated in~\cite{adams2012} that we do not duplicate here. 

%
%
%
%
\section{Discussion and conclusions}
\label{sec:conclusions}
\noindent

We have presented an analysis framework, which can be used to extract information about the properties of the Milky Way by characterizing the spectral shape or the residual foreground signal of the DWD binaries measured by \textit{LISA}. Based on a fiducial binary population synthesis model, we first began by simulating catalogues of sources using different parameters describing the total stellar mass and the shape of the Galaxy. For each of them, we estimated the residual foreground signal after subtracting the `loudest' sources, based on SNR criteria following the methodology presented in~\citet{Karnesis2021tsh, nis12, Timpano2006, crowder2007} (see~\cref{sec:foreground_estimation} for details).

As our first investigation, we designed an analysis framework to estimate the overall Galactic mass locked in stars from the properties of the stochastic GW signal. We used an empirical model for the spectral shape of the confusion signal, and then related its parameters to the total number of sources produced by our fiducial population of Galactic DWDs. In turn, we can relate the number of sources to the Galactic stellar mass by reverse engineering our synthetic models \citep{Korol2021}. This is possible as the assumptions on the Galactic star formation history and the current age set the total stellar mass of the Galaxy, while the DWD evolution model sets the number of DWDs in the \textit{LISA} frequency band per unit stellar mass at any given time. Thus, if these assumptions are kept fixed, we can recover the stellar mass necessary to produce the `observed' spectral shape of the confusion signal. 

To test this idea, we generated different catalogues by randomly removing entries in proportion to the fraction of the total number of DWD in our fiducial catalogue $N_{\%} = \{ 0.5, 0.6, 0.7, 0.8, 0.9, 1\}$, which is in turn proportional to the total stellar mass assumed in our model $m_{\mathrm{G}} \equiv 8.2 \times 10^{10}\,$M$\odot$. We then estimated the residual foreground stochastic signal for these catalogues. Using an empirical model for the spectral shape of the signal, we associated its parameters to the total number of sources per catalogue. This was done by performing a joint fit on the equivalent datasets and producing the linear relations that are shown in~\cref{fig:mass_params}. With these relations assumed known, we reversed the argument to compute the errors for the different cases for the Galactic mass considered, which are presented in~\cref{tab:npercent}. The estimated uncertainty is, as expected, quite small. This can be attributed to our assumption of complete knowledge of the underlying population model of the DWD binaries. However, we expect that as we relax some of our assumptions on the model parameters, the estimated error bars will increase.

Uncertainties of the Milky Way's total stellar mass estimates based on electromagnetic observations are of the order of $\sim 10$\,per cent \citep[e.g.][and references therein]{lic15,bla16,cau20}. As in our case, these estimates are model-dependent relying on a number of similar assumptions, such as the initial mass function, star formation history and the density profiles of the Galactic stellar components. However, the framework developed here constitutes an independent and alternative approach. While electromagnetic estimates are typically based on bright young stars, we have proposed to use GWs emitted by DWDs that are inaccessible to electromagnetic observatories at distances larger than a few kpc. We also stress that GW detections yield the original stellar mass of the Galaxy, i.e. including the contribution from evolved stellar populations that may not be visible. This is in contrast to masses derived from electromagnetic brightness, sensitive to the mass enclosed in bright stars, which are made by modeling the brightness of a population and applying age-dependent luminosity-to-mass ratios from stellar calculations.

We highlight that our analysis is conducted for a particular binary population model. Thus, one should expect the results of the fit to the shape of the  stochastic foreground to change based on a different model; the concept explored here would still hold. We also note that we considered some of the parameters regulating the total number of DWDs in the catalogue fixed. For instance, the initial binary fraction is part of the initial conditions that regulates the number of binaries produced by a DWD evolution model. As already discussed, it is effectively degenerate to the total stellar mass. While it could be adjusted in post-processing or estimated from electromagnetic observations, it is true that our analysis of the total Galactic mass is limited by its assumed value. Or in other words, inference of the Galaxy mass will only be as precise as our knowledge of the initial and final (i.e. DWD) binary fraction, which are becoming more precise as observation samples grow~\citep{Duchene2013,Bad18,belokurov2020, Korol2020,maozhallakoun,napi}. We also highlight that the fiducial DWD model employed for our study matches well the DWD fraction of the local {\it Gaia} sample \citep{Toonen2017}.. The assumption on the binary fraction also regulates the level of the confusion foreground similarly to the total stellar mass \citep[e.g.][]{thi21}. In addition, assumptions regarding physical processes involved in DWD evolution, such as the common envelope efficiency, also impact the number and the properties of DWDs. For simplicity, here we chose to fix all the parameters entering the DWD evolution model and vary only those characterizing our Galactic model. We defer further complementary investigations to future work.

Next, motivated by existing studies, we focused on probing the scale height of the Galaxy by measuring the stochastic foreground signal. We found that for values of $Z_\mathrm{d} \lesssim 600~\mathrm{pc}$, \textit{LISA} will not be able to distinguish between the different cases based on measurements of the stochastic signal PSD shape alone. We confirmed those findings by comparing the total number and SNR values of the resolved sources, as well as the relative uncertainties of the waveform parameters. This is in line with the previous work of~\cite{Breivik2020}, in which authors using spherical harmonics analysis, have shown that differences of scale height with $Z_\mathrm{d} < 600~\mathrm{pc}$ are extremely challenging to distinguish by the confusion signal alone.
In more details, based on studies such as \cite{Benacquista2006}, if the local DWD space density is considered constant ($\rho_{\rm DWD}$), then an increase in the scale height implies an increase in the volume of the Galaxy and, consequently, in total number of DWDs populating it, \emph{i.e.} $N_{\rm DWD} = \rho_{\rm DWD} V$, with $V$ being the result of the integration of equations \eqref{eqn:disc} and \eqref{eqn:bulge}. Therefore, both the overall strength of the confusion-limited signal and the transition frequency -- the frequency where most signals become individually resolvable -- increases. On the other hand, if the total number of binaries $N_{\rm DWD}$ is held constant, then an increase in the scale height leads to a reduction of the number of resolvable signals above the confusion-limited signal at low frequencies. Finally, \citet{Benacquista2006} conclude that if the number of the sources is fixed by a global calibration method, changing the scale height of the Galaxy, which implies a decrease in the DWD binaries density to keep $N$ constant) does not have a major effect on the results. In addition, in this latter case, as the number $N_{\rm DWD}$ is independent of the scale height, the expected transition frequency  will remain the same. From  \cref{fig:Z_5} it becomes apparent that the three catalogues using $Z_\mathrm{d} = \{100,\, 300,\, 500\}~\mathrm{pc}$ yield similar unresolved signals, in agreement with the results  obtained by \cite{Benacquista2006}. According to the right panel of~\cref{fig:Z_5},~\cref{fig:zfisher},  and~\cref{tab:zsources}, the total number of resolved sources, as well as their corresponding SNR and estimated waveform parameter uncertainties, is very similar for all cases. This means that, for the cases of $Z_\mathrm{d}$ considered, the disc scale height does not significantly impact the resolvability of `louder' sources. 

Finally, we defined a hierarchical Bayesian model, which was used to probe the characteristics of interesting population parameters, namely the chirp mass ${\mathcal M}$, the $f_0$ and the time of coalescence $\tau_c$ distributions. The distributions of these three quantities measured by \textit{LISA} can be directly compared with electromagnetic observations of DWDs and will be fundamental for understanding the evolution of the DWDs systems and the supernova type Ia progenitors. In the same spirit as in~\cite{adams2012}, we used the catalogue of resolved sources, and their corresponding estimate of their waveform uncertainties (calculated from the Fisher Information Matrices), in order to reconstruct their hyper-prior densities. We recover the densities of $f_0$ and $\tau_c$ with very good accuracy and with no bias, while the chirp-mass density is inaccessible for values of ${\mathcal M}\gtrapprox 0.7~\mathrm{M}_\odot$. This can be attributed to the selection effects of the sources, and is evident from the right panel of~\cref{fig:pdfs}.  

In this work, we have presented a number of investigations that will be possible to be addressed with  \textit{LISA} data, focusing on DWDs. Many more open questions are still awaiting to be explored and/or quantitatively assessed ahead of  \textit{LISA}'s launch \citep{amaro22}.
In future work, we plan a joint analysis of resolved and unresolved sources, under the same Bayesian framework. While in this work we use a fiducial DWD model, changing only those parameters describing the mass and the shape of the Galaxy, our preliminary investigations show that changes to binary evolution also produces changes to the level and the shape of the Galactic unresolved foreground. Including alternative DWD models will also make our estimates on parameters' uncertainties more realistic.   

\section*{Acknowledgements}
We thank Silvia Toonen for providing us with a DWD evolution model used to assemble our fiducial  Galactic population.
NS and NK acknowledge  support from the Gr-PRODEX 2019 funding program (PEA 4000132310), and from the {\it SpaceSHEL} project funded by the  Hellenic Foundation for Research \& Innovation. 
VK acknowledges support from the Netherlands Research Council NWO (Rubicon 019.183EN.015 grant).
The work of M.P. was supported by STFC grants ST/P000762/1 and ST/T000791/1. M.P. acknowledges support by the European Unions' Horizon 2020 Research Council grant 724659 MassiveCosmo ERC- 2016-COG.\\
This research made use of the tools provided by the \textit{LISA} Data Processing Group (LDPG) and the \textit{LISA} Consortium \textit{LISA} Data Challenges (LDC) working group\footnote{\href{https://lisa-ldc.lal.in2p3.fr/}{https://lisa-ldc.lal.in2p3.fr/}}.

\section*{Data Availability}
\label{sec:data_availability}
\noindent
The data and codes used in the analysis are publicly available in the following \texttt{github} repository: \url{https://gitlab.in2p3.fr/Nikos/galactic_properties_from_cgbs}. For sampling the posterior surfaces, we used a Metropolis-Hastings MCMC, which can be accessed at  
\url{https://gitlab.in2p3.fr/Nikos/metropolishastings.git}. The sampler employs  Simulated Annealing and adapting proposal distribution techniques. To estimate the residual foreground signal of the DWD binaries, we used tools from the LDC software (\url{https://lisa-ldc.lal.in2p3.fr/code}) and the \textsc{GWG} software package  \url{https://gitlab.in2p3.fr/Nikos/gwg}. 



\bibliographystyle{mnras}

\begin{thebibliography}{}
\makeatletter
\relax
\def\mn@urlcharsother{\let\do\@makeother \do\$\do\&\do\#\do\^\do\_\do\%\do\~}
\def\mn@doi{\begingroup\mn@urlcharsother \@ifnextchar [ {\mn@doi@}
  {\mn@doi@[]}}
\def\mn@doi@[#1]#2{\def\@tempa{#1}\ifx\@tempa\@empty \href
  {http://dx.doi.org/#2} {doi:#2}\else \href {http://dx.doi.org/#2} {#1}\fi
  \endgroup}
\def\mn@eprint#1#2{\mn@eprint@#1:#2::\@nil}
\def\mn@eprint@arXiv#1{\href {http://arxiv.org/abs/#1} {{\tt arXiv:#1}}}
\def\mn@eprint@dblp#1{\href {http://dblp.uni-trier.de/rec/bibtex/#1.xml}
  {dblp:#1}}
\def\mn@eprint@#1:#2:#3:#4\@nil{\def\@tempa {#1}\def\@tempb {#2}\def\@tempc
  {#3}\ifx \@tempc \@empty \let \@tempc \@tempb \let \@tempb \@tempa \fi \ifx
  \@tempb \@empty \def\@tempb {arXiv}\fi \@ifundefined
  {mn@eprint@\@tempb}{\@tempb:\@tempc}{\expandafter \expandafter \csname
  mn@eprint@\@tempb\endcsname \expandafter{\@tempc}}}

\bibitem[\protect\citeauthoryear{Abbott et~al.}{Abbott
  et~al.}{2021}]{LIGOScientific:2020kqk}
Abbott R.,  et~al., 2021, \mn@doi [Astrophys. J. Lett.]
  {10.3847/2041-8213/abe949}, 913, L7

\bibitem[\protect\citeauthoryear{{Abt}}{{Abt}}{1983}]{Abt1983}
{Abt} H.~A.,  1983, \mn@doi [\araa] {10.1146/annurev.aa.21.090183.002015},
  \href {http://adsabs.harvard.edu/abs/1983ARA\%26A..21..343A} {21, 343}

\bibitem[\protect\citeauthoryear{Adams, Cornish  \& Littenberg}{Adams
  et~al.}{2012}]{adams2012}
Adams M.~R.,  Cornish N.~J.,   Littenberg T.~B.,  2012, \mn@doi [Phys. Rev. D]
  {10.1103/PhysRevD.86.124032}, 86, 124032

\bibitem[\protect\citeauthoryear{{Amaro-Seoane} et~al.,}{{Amaro-Seoane}
  et~al.}{2017}]{LISAwhitepaper}
{Amaro-Seoane} P.,  et~al., 2017, arXiv e-prints, \href
  {https://ui.adsabs.harvard.edu/abs/2017arXiv170200786A} {p. arXiv:1702.00786}

\bibitem[\protect\citeauthoryear{{Amaro-Seoane} et~al.,}{{Amaro-Seoane}
  et~al.}{2022}]{amaro22}
{Amaro-Seoane} P.,  et~al., 2022, arXiv e-prints, \href
  {https://ui.adsabs.harvard.edu/abs/2022arXiv220306016A} {p. arXiv:2203.06016}

\bibitem[\protect\citeauthoryear{{Artale}, {Zehavi}, {Contreras}  \&
  {Norberg}}{{Artale} et~al.}{2018}]{art2018}
{Artale} M.~C.,  {Zehavi} I.,  {Contreras} S.,   {Norberg} P.,  2018, \mn@doi
  [\mnras] {10.1093/mnras/sty2110}, \href
  {https://ui.adsabs.harvard.edu/abs/2018MNRAS.480.3978A} {480, 3978}

\bibitem[\protect\citeauthoryear{{Babak} et~al.,}{{Babak}
  et~al.}{2017}]{Babak2017}
{Babak} S.,  et~al., 2017, \prd, 95, 103012

\bibitem[\protect\citeauthoryear{Babak, Petiteau  \& Hewitson}{Babak
  et~al.}{2021}]{Babak2021mhe}
Babak S.,  Petiteau A.,   Hewitson M.,  2021

\bibitem[\protect\citeauthoryear{{Badenes} et~al.,}{{Badenes}
  et~al.}{2018}]{Bad18}
{Badenes} C.,  et~al., 2018, \mn@doi [\apj] {10.3847/1538-4357/aaa765}, \href
  {https://ui.adsabs.harvard.edu/abs/2018ApJ...854..147B} {854, 147}

\bibitem[\protect\citeauthoryear{Baghi, Thorpe, Slutsky, Baker, Dal~Canton,
  Korsakova  \& Karnesis}{Baghi et~al.}{2019}]{Baghi19}
Baghi Q.,  Thorpe I.,  Slutsky J.,  Baker J.,  Dal~Canton T.,  Korsakova N.,
  Karnesis N.,  2019, \mn@doi [Phys. Rev. D] {10.1103/PhysRevD.100.022003},
  100, 022003

\bibitem[\protect\citeauthoryear{{Belokurov} et~al.,}{{Belokurov}
  et~al.}{2020}]{belokurov2020}
{Belokurov} V.,  et~al., 2020, \mn@doi [\mnras] {10.1093/mnras/staa1522}, \href
  {https://ui.adsabs.harvard.edu/abs/2020MNRAS.496.1922B} {496, 1922}

\bibitem[\protect\citeauthoryear{{Benacquista} \&
  {Holley-Bockelmann}}{{Benacquista} \&
  {Holley-Bockelmann}}{2006}]{Benacquista2006}
{Benacquista} M.,  {Holley-Bockelmann} K.,  2006, \mn@doi [\apj]
  {10.1086/504024}, \href
  {https://ui.adsabs.harvard.edu/abs/2006ApJ...645..589B} {645, 589}

\bibitem[\protect\citeauthoryear{Bender \& Hils}{Bender \&
  Hils}{1997}]{Bender1997hs}
Bender P.~L.,  Hils D.,  1997, \mn@doi [Class. Quant. Grav.]
  {10.1088/0264-9381/14/6/008}, 14, 1439

\bibitem[\protect\citeauthoryear{{Bland-Hawthorn} \&
  {Gerhard}}{{Bland-Hawthorn} \& {Gerhard}}{2016}]{bla16}
{Bland-Hawthorn} J.,  {Gerhard} O.,  2016, \mn@doi [\araa]
  {10.1146/annurev-astro-081915-023441}, \href
  {https://ui.adsabs.harvard.edu/abs/2016ARA&A..54..529B} {54, 529}

\bibitem[\protect\citeauthoryear{B\l{}aut, Babak  \& Kr\'olak}{B\l{}aut
  et~al.}{2010}]{blaut}
B\l{}aut A.,  Babak S.,   Kr\'olak A.,  2010, \mn@doi [Phys. Rev. D]
  {10.1103/PhysRevD.81.063008}, 81, 063008

\bibitem[\protect\citeauthoryear{Boileau, Lamberts, Cornish  \& Meyer}{Boileau
  et~al.}{2021}]{Boileau2021}
Boileau G.,  Lamberts A.,  Cornish N.~J.,   Meyer R.,  2021, \mn@doi [Mon. Not.
  Roy. Astron. Soc.] {10.1093/mnras/stab2575}, 508, 803

\bibitem[\protect\citeauthoryear{{Boissier} \& {Prantzos}}{{Boissier} \&
  {Prantzos}}{1999}]{BP99}
{Boissier} S.,  {Prantzos} N.,  1999, \mn@doi [\mnras]
  {10.1046/j.1365-8711.1999.02699.x}, \href
  {https://ui.adsabs.harvard.edu/abs/1999MNRAS.307..857B} {307, 857}

\bibitem[\protect\citeauthoryear{{Bonetti} \& {Sesana}}{{Bonetti} \&
  {Sesana}}{2020}]{bon20}
{Bonetti} M.,  {Sesana} A.,  2020, \mn@doi [\prd]
  {10.1103/PhysRevD.102.103023}, \href
  {https://ui.adsabs.harvard.edu/abs/2020PhRvD.102j3023B} {102, 103023}

\bibitem[\protect\citeauthoryear{{Breivik} et~al.,}{{Breivik}
  et~al.}{2020a}]{bre20}
{Breivik} K.,  et~al., 2020a, \mn@doi [\apj] {10.3847/1538-4357/ab9d85}, \href
  {https://ui.adsabs.harvard.edu/abs/2020ApJ...898...71B} {898, 71}

\bibitem[\protect\citeauthoryear{{Breivik}, {Mingarelli}  \&
  {Larson}}{{Breivik} et~al.}{2020b}]{Breivik2020}
{Breivik} K.,  {Mingarelli} C. M.~F.,   {Larson} S.~L.,  2020b, \mn@doi [\apj]
  {10.3847/1538-4357/abab99}, \href
  {https://ui.adsabs.harvard.edu/abs/2020ApJ...901....4B} {901, 4}

\bibitem[\protect\citeauthoryear{{Bruzual A.}}{{Bruzual A.}}{1983}]{bru83}
{Bruzual A.} G.,  1983, \mn@doi [\apj] {10.1086/161352}, \href
  {https://ui.adsabs.harvard.edu/abs/1983ApJ...273..105B} {273, 105}

\bibitem[\protect\citeauthoryear{Caprini \& Figueroa}{Caprini \&
  Figueroa}{2018}]{Caprini2018mtu}
Caprini C.,  Figueroa D.~G.,  2018, \mn@doi [Class. Quant. Grav.]
  {10.1088/1361-6382/aac608}, 35, 163001

\bibitem[\protect\citeauthoryear{{Caprini} et~al.,}{{Caprini}
  et~al.}{2016}]{Caprini2016}
{Caprini} C.,  et~al., 2016, \jcap, 4, 001

\bibitem[\protect\citeauthoryear{Caprini, Figueroa, Flauger, Nardini, Peloso,
  Pieroni, Ricciardone  \& Tasinato}{Caprini et~al.}{2019}]{Caprini:2019pxz}
Caprini C.,  Figueroa D.~G.,  Flauger R.,  Nardini G.,  Peloso M.,  Pieroni M.,
   Ricciardone A.,   Tasinato G.,  2019, \mn@doi [JCAP]
  {10.1088/1475-7516/2019/11/017}, 11, 017

\bibitem[\protect\citeauthoryear{{Cautun} et~al.,}{{Cautun}
  et~al.}{2020}]{cau20}
{Cautun} M.,  et~al., 2020, \mn@doi [\mnras] {10.1093/mnras/staa1017}, \href
  {https://ui.adsabs.harvard.edu/abs/2020MNRAS.494.4291C} {494, 4291}

\bibitem[\protect\citeauthoryear{Cornish \& Littenberg}{Cornish \&
  Littenberg}{2007}]{corn07}
Cornish N.~J.,  Littenberg T.~B.,  2007, \mn@doi [Phys. Rev. D]
  {10.1103/PhysRevD.76.083006}, 76, 083006

\bibitem[\protect\citeauthoryear{Crowder \& Cornish}{Crowder \&
  Cornish}{2007}]{crowder2007}
Crowder J.,  Cornish N.~J.,  2007, \mn@doi [Phys. Rev. D]
  {10.1103/PhysRevD.75.043008}, 75, 043008

\bibitem[\protect\citeauthoryear{{Dayal}, {Rossi}, {Shiralilou}, {Piana},
  {Choudhury}  \& {Volonteri}}{{Dayal} et~al.}{2019}]{Dayal2019}
{Dayal} P.,  {Rossi} E.~M.,  {Shiralilou} B.,  {Piana} O.,  {Choudhury} T.~R.,
   {Volonteri} M.,  2019, \mn@doi [\mnras] {10.1093/mnras/stz897}, \href
  {https://ui.adsabs.harvard.edu/abs/2019MNRAS.486.2336D} {486, 2336}

\bibitem[\protect\citeauthoryear{Dey, Karnesis, Toubiana, Barausse, Korsakova,
  Baghi  \& Basak}{Dey et~al.}{2021}]{dey21}
Dey K.,  Karnesis N.,  Toubiana A.,  Barausse E.,  Korsakova N.,  Baghi Q.,
  Basak S.,  2021, \mn@doi [Phys. Rev. D] {10.1103/PhysRevD.104.044035}, 104,
  044035

\bibitem[\protect\citeauthoryear{Digman \& Cornish}{Digman \&
  Cornish}{2022}]{Digman2022jmp}
Digman M.~C.,  Cornish N.~J.,  2022

\bibitem[\protect\citeauthoryear{{Duch{\^e}ne} \& {Kraus}}{{Duch{\^e}ne} \&
  {Kraus}}{2013}]{Duchene2013}
{Duch{\^e}ne} G.,  {Kraus} A.,  2013, \mn@doi [\araa]
  {10.1146/annurev-astro-081710-102602}, \href
  {https://ui.adsabs.harvard.edu/abs/2013ARA&A..51..269D} {51, 269}

\bibitem[\protect\citeauthoryear{{El-Badry} \& {Rix}}{{El-Badry} \&
  {Rix}}{2018}]{El-Badry2018}
{El-Badry} K.,  {Rix} H.-W.,  2018, \mn@doi [\mnras] {10.1093/mnras/sty2186},
  \href {https://ui.adsabs.harvard.edu/abs/2018MNRAS.480.4884E} {480, 4884}

\bibitem[\protect\citeauthoryear{{El-Badry} \& {Rix}}{{El-Badry} \&
  {Rix}}{2019}]{Elb19}
{El-Badry} K.,  {Rix} H.-W.,  2019, \mn@doi [\mnras] {10.1093/mnrasl/sly206},
  \href {https://ui.adsabs.harvard.edu/abs/2019MNRAS.482L.139E} {482, L139}

\bibitem[\protect\citeauthoryear{{Fantin} et~al.,}{{Fantin}
  et~al.}{2019}]{Fantin2019}
{Fantin} N.~J.,  et~al., 2019, \mn@doi [\apj] {10.3847/1538-4357/ab5521}, \href
  {https://ui.adsabs.harvard.edu/abs/2019ApJ...887..148F} {887, 148}

\bibitem[\protect\citeauthoryear{Flauger, Karnesis, Nardini, Pieroni,
  Ricciardone  \& Torrado}{Flauger et~al.}{2021}]{Flauger:2020qyi}
Flauger R.,  Karnesis N.,  Nardini G.,  Pieroni M.,  Ricciardone A.,   Torrado
  J.,  2021, \mn@doi [JCAP] {10.1088/1475-7516/2021/01/059}, 01, 059

\bibitem[\protect\citeauthoryear{Gelman, Carlin, Stern  \& Rubin}{Gelman
  et~al.}{2004}]{gelmanbda04}
Gelman A.,  Carlin J.~B.,  Stern H.~S.,   Rubin D.~B.,  2004, Bayesian Data
  Analysis, 2nd ed. edn.
Chapman and Hall/CRC

\bibitem[\protect\citeauthoryear{Georgousi}{Georgousi}{2021}]{marythesis}
Georgousi M.,  2021,
  \href{https://ikee.lib.auth.gr/record/338425/files/GEORGOUSI.pdf}{Probing the
  Properties of our Galaxy by Detecting Gravitational Waves from Ultra Compact
  Binaries with LISA}

\bibitem[\protect\citeauthoryear{{Gravity Collaboration} et~al.,}{{Gravity
  Collaboration} et~al.}{2019}]{abu19}
{Gravity Collaboration} et~al., 2019, \mn@doi [\aap]
  {10.1051/0004-6361/201935656}, \href
  {https://ui.adsabs.harvard.edu/abs/2019A&A...625L..10G} {625, L10}

\bibitem[\protect\citeauthoryear{Green}{Green}{1995}]{rjmcmc}
Green P.~J.,  1995, Biometrika, 82, 711

\bibitem[\protect\citeauthoryear{{Heggie}}{{Heggie}}{1975}]{Heggie1975}
{Heggie} D.~C.,  1975, \mn@doi [\mnras] {10.1093/mnras/173.3.729}, \href
  {https://ui.adsabs.harvard.edu/abs/1975MNRAS.173..729H} {173, 729}

\bibitem[\protect\citeauthoryear{Huang et~al.,}{Huang et~al.}{2020}]{tianquin}
Huang S.-J.,  et~al., 2020, \mn@doi [Phys. Rev. D]
  {10.1103/PhysRevD.102.063021}, 102, 063021

\bibitem[\protect\citeauthoryear{{Ivanova} et~al.,}{{Ivanova}
  et~al.}{2013}]{Ivanova2013}
{Ivanova} N.,  et~al., 2013, \mn@doi [\aapr] {10.1007/s00159-013-0059-2}, \href
  {http://adsabs.harvard.edu/abs/2013A%26ARv..21...59I} {21, 59}

\bibitem[\protect\citeauthoryear{{Juri{\'c}} et~al.,}{{Juri{\'c}}
  et~al.}{2008}]{Juric2008}
{Juri{\'c}} M.,  et~al., 2008, \apj, 673, 864

\bibitem[\protect\citeauthoryear{Karnesis, Babak, Pieroni, Cornish  \&
  Littenberg}{Karnesis et~al.}{2021}]{Karnesis2021tsh}
Karnesis N.,  Babak S.,  Pieroni M.,  Cornish N.,   Littenberg T.,  2021,
  \mn@doi [Phys. Rev. D] {10.1103/PhysRevD.104.043019}, 104, 043019

\bibitem[\protect\citeauthoryear{{Klein} et~al.,}{{Klein}
  et~al.}{2016}]{Klein2016}
{Klein} A.,  et~al., 2016, \prd, 93, 024003

\bibitem[\protect\citeauthoryear{{Korol}, {Rossi}, {Groot}, {Nelemans},
  {Toonen}  \& {Brown}}{{Korol} et~al.}{2017}]{Korol2017}
{Korol} V.,  {Rossi} E.~M.,  {Groot} P.~J.,  {Nelemans} G.,  {Toonen} S.,
  {Brown} A. G.~A.,  2017, \mn@doi [\mnras] {10.1093/mnras/stx1285}, \href
  {https://ui.adsabs.harvard.edu/abs/2017MNRAS.470.1894K} {470, 1894}

\bibitem[\protect\citeauthoryear{{Korol}, {Rossi}  \& {Barausse}}{{Korol}
  et~al.}{2019}]{Korol2019}
{Korol} V.,  {Rossi} E.~M.,   {Barausse} E.,  2019, \mn@doi [\mnras]
  {10.1093/mnras/sty3440}, \href
  {https://ui.adsabs.harvard.edu/abs/2019MNRAS.483.5518K} {483, 5518}

\bibitem[\protect\citeauthoryear{{Korol} et~al.,}{{Korol}
  et~al.}{2020}]{Korol2020}
{Korol} V.,  et~al., 2020, \mn@doi [\aap] {10.1051/0004-6361/202037764}, \href
  {https://ui.adsabs.harvard.edu/abs/2020A&A...638A.153K} {638, A153}

\bibitem[\protect\citeauthoryear{{Korol}, {Belokurov}, {Moore}  \&
  {Toonen}}{{Korol} et~al.}{2021}]{Korol2021}
{Korol} V.,  {Belokurov} V.,  {Moore} C.~J.,   {Toonen} S.,  2021, \mn@doi
  [\mnras] {10.1093/mnrasl/slab003}, \href
  {https://ui.adsabs.harvard.edu/abs/2021MNRAS.502L..55K} {502, L55}

\bibitem[\protect\citeauthoryear{{Korol}, {Hallakoun}, {Toonen}  \&
  {Karnesis}}{{Korol} et~al.}{2022}]{KorolHallakoun2021}
{Korol} V.,  {Hallakoun} N.,  {Toonen} S.,   {Karnesis} N.,  2022, \mn@doi
  [\mnras] {10.1093/mnras/stac415}, \href
  {https://ui.adsabs.harvard.edu/abs/2022MNRAS.511.5936K} {511, 5936}

\bibitem[\protect\citeauthoryear{{Kroupa}, {Tout}  \& {Gilmore}}{{Kroupa}
  et~al.}{1993}]{KroupaIMF}
{Kroupa} P.,  {Tout} C.~A.,   {Gilmore} G.,  1993, \mn@doi [\mnras]
  {10.1093/mnras/262.3.545}, \href
  {http://adsabs.harvard.edu/abs/1993MNRAS.262..545K} {262, 545}

\bibitem[\protect\citeauthoryear{{Kupfer} et~al.,}{{Kupfer}
  et~al.}{2018}]{Kupfer2018}
{Kupfer} T.,  et~al., 2018, \mn@doi [\mnras] {10.1093/mnras/sty1545}, \href
  {https://ui.adsabs.harvard.edu/abs/2018MNRAS.480..302K} {480, 302}

\bibitem[\protect\citeauthoryear{{LISA Science Study Team}}{{LISA Science Study
  Team}}{2018}]{SciRD}
{LISA Science Study Team} 2018, Technical Report~1.0, {LISA Science
  Requirements Document, ESA-L3-EST-SCI-RS-001}.
ESA

\bibitem[\protect\citeauthoryear{{Lamberts} et~al.,}{{Lamberts}
  et~al.}{2018}]{lam18}
{Lamberts} A.,  et~al., 2018, \mn@doi [\mnras] {10.1093/mnras/sty2035}, \href
  {https://ui.adsabs.harvard.edu/abs/2018MNRAS.480.2704L} {480, 2704}

\bibitem[\protect\citeauthoryear{{Lamberts}, {Blunt}, {Littenberg},
  {Garrison-Kimmel}, {Kupfer}  \& {Sanderson}}{{Lamberts} et~al.}{2019}]{lam19}
{Lamberts} A.,  {Blunt} S.,  {Littenberg} T.~B.,  {Garrison-Kimmel} S.,
  {Kupfer} T.,   {Sanderson} R.~E.,  2019, \mn@doi [\mnras]
  {10.1093/mnras/stz2834}, \href
  {https://ui.adsabs.harvard.edu/abs/2019MNRAS.490.5888L} {490, 5888}

\bibitem[\protect\citeauthoryear{{Li}, {Chen}, {Chen}, {Li}, {Yu}  \&
  {Han}}{{Li} et~al.}{2020}]{zen20}
{Li} Z.,  {Chen} X.,  {Chen} H.-L.,  {Li} J.,  {Yu} S.,   {Han} Z.,  2020,
  \mn@doi [\apj] {10.3847/1538-4357/ab7dc2}, \href
  {https://ui.adsabs.harvard.edu/abs/2020ApJ...893....2L} {893, 2}

\bibitem[\protect\citeauthoryear{{Licquia} \& {Newman}}{{Licquia} \&
  {Newman}}{2015}]{lic15}
{Licquia} T.~C.,  {Newman} J.~A.,  2015, \mn@doi [\apj]
  {10.1088/0004-637X/806/1/96}, \href
  {https://ui.adsabs.harvard.edu/abs/2015ApJ...806...96L} {806, 96}

\bibitem[\protect\citeauthoryear{Littenberg, Cornish, Lackeos  \&
  Robson}{Littenberg et~al.}{2020}]{Litten2020}
Littenberg T.,  Cornish N.,  Lackeos K.,   Robson T.,  2020, \mn@doi [Phys.
  Rev. D] {10.1103/PhysRevD.101.123021}, 101, 123021

\bibitem[\protect\citeauthoryear{{Mackereth} et~al.,}{{Mackereth}
  et~al.}{2017}]{mac17}
{Mackereth} J.~T.,  et~al., 2017, \mn@doi [\mnras] {10.1093/mnras/stx1774},
  \href {https://ui.adsabs.harvard.edu/abs/2017MNRAS.471.3057M} {471, 3057}

\bibitem[\protect\citeauthoryear{Maggiore}{Maggiore}{2007}]{Maggiore}
Maggiore M.,  2007, {Gravitational Waves. Vol. 1: Theory and Experiments}.
Oxford Master Series in Physics, Oxford University Press

\bibitem[\protect\citeauthoryear{{Maoz} \& {Hallakoun}}{{Maoz} \&
  {Hallakoun}}{2017}]{maozhallakoun}
{Maoz} D.,  {Hallakoun} N.,  2017, \mn@doi [\mnras] {10.1093/mnras/stx102},
  \href {https://ui.adsabs.harvard.edu/abs/2017MNRAS.467.1414M} {467, 1414}

\bibitem[\protect\citeauthoryear{{Maoz} \& {Mannucci}}{{Maoz} \&
  {Mannucci}}{2012}]{Maoz2012}
{Maoz} D.,  {Mannucci} F.,  2012, \mn@doi [\pasa] {10.1071/AS11052}, \href
  {https://ui.adsabs.harvard.edu/abs/2012PASA...29..447M} {29, 447}

\bibitem[\protect\citeauthoryear{{Maraston}}{{Maraston}}{1998}]{mar98}
{Maraston} C.,  1998, \mn@doi [\mnras] {10.1046/j.1365-8711.1998.01947.x},
  \href {https://ui.adsabs.harvard.edu/abs/1998MNRAS.300..872M} {300, 872}

\bibitem[\protect\citeauthoryear{{Moe}, {Kratter}  \& {Badenes}}{{Moe}
  et~al.}{2019}]{Moe19}
{Moe} M.,  {Kratter} K.~M.,   {Badenes} C.,  2019, \mn@doi [\apj]
  {10.3847/1538-4357/ab0d88}, \href
  {https://ui.adsabs.harvard.edu/abs/2019ApJ...875...61M} {875, 61}

\bibitem[\protect\citeauthoryear{{Moore}, {Chua}  \& {Gair}}{{Moore}
  et~al.}{2017}]{moo17}
{Moore} C.~J.,  {Chua} A. J.~K.,   {Gair} J.~R.,  2017, \mn@doi [Classical and
  Quantum Gravity] {10.1088/1361-6382/aa85fa}, \href
  {https://ui.adsabs.harvard.edu/abs/2017CQGra..34s5009M} {34, 195009}

\bibitem[\protect\citeauthoryear{{Napiwotzki} et~al.,}{{Napiwotzki}
  et~al.}{2020}]{napi}
{Napiwotzki} R.,  et~al., 2020, \mn@doi [\aap] {10.1051/0004-6361/201629648},
  \href {https://ui.adsabs.harvard.edu/abs/2020A&A...638A.131N} {638, A131}

\bibitem[\protect\citeauthoryear{{Nelemans} \& {Tout}}{{Nelemans} \&
  {Tout}}{2005}]{nel05}
{Nelemans} G.,  {Tout} C.~A.,  2005, \mn@doi [\mnras]
  {10.1111/j.1365-2966.2004.08496.x}, \href
  {https://ui.adsabs.harvard.edu/abs/2005MNRAS.356..753N} {356, 753}

\bibitem[\protect\citeauthoryear{{Nelemans}, {Verbunt}, {Yungelson}  \&
  {Portegies Zwart}}{{Nelemans} et~al.}{2000}]{Nelemans2000}
{Nelemans} G.,  {Verbunt} F.,  {Yungelson} L.~R.,   {Portegies Zwart} S.~F.,
  2000, \aap, 360, 1011

\bibitem[\protect\citeauthoryear{{Nelemans}, {Yungelson}, {Portegies Zwart}  \&
  {Verbunt}}{{Nelemans} et~al.}{2001}]{Nelemans01a}
{Nelemans} G.,  {Yungelson} L.~R.,  {Portegies Zwart} S.~F.,   {Verbunt} F.,
  2001, \mn@doi [\aap] {10.1051/0004-6361:20000147}, \href
  {https://ui.adsabs.harvard.edu/abs/2001A&A...365..491N} {365, 491}

\bibitem[\protect\citeauthoryear{{Nelemans}, {Yungelson}  \& {Portegies
  Zwart}}{{Nelemans} et~al.}{2004}]{Nelemans2004}
{Nelemans} G.,  {Yungelson} L.~R.,   {Portegies Zwart} S.~F.,  2004, \mn@doi
  [\mnras] {10.1111/j.1365-2966.2004.07479.x}, \href
  {https://ui.adsabs.harvard.edu/abs/2004MNRAS.349..181N} {349, 181}

\bibitem[\protect\citeauthoryear{Nesti \& Salucci}{Nesti \&
  Salucci}{2013}]{Nesti_2013}
Nesti F.,  Salucci P.,  2013, \mn@doi [Journal of Cosmology and Astroparticle
  Physics] {10.1088/1475-7516/2013/07/016}, 2013, 016

\bibitem[\protect\citeauthoryear{{Nissanke}, {Vallisneri}, {Nelemans}  \&
  {Prince}}{{Nissanke} et~al.}{2012}]{nis12}
{Nissanke} S.,  {Vallisneri} M.,  {Nelemans} G.,   {Prince} T.~A.,  2012,
  \mn@doi [\apj] {10.1088/0004-637X/758/2/131}, \href
  {https://ui.adsabs.harvard.edu/abs/2012ApJ...758..131N} {758, 131}

\bibitem[\protect\citeauthoryear{O'Brien, Szczepa\ifmmode~\acute{n}\else
  \'{n}\fi{}czyk, Gayathri, Bartos, Vedovato, Prodi, Mitselmakher  \&
  Klimenko}{O'Brien et~al.}{2021}]{PhysRevD.104.082003}
O'Brien B.,  Szczepa\ifmmode~\acute{n}\else \'{n}\fi{}czyk M.,  Gayathri V.,
  Bartos I.,  Vedovato G.,  Prodi G.,  Mitselmakher G.,   Klimenko S.,  2021,
  \mn@doi [Phys. Rev. D] {10.1103/PhysRevD.104.082003}, 104, 082003

\bibitem[\protect\citeauthoryear{{Paczynski}}{{Paczynski}}{1976}]{Paczynski1976}
{Paczynski} B.,  1976, in {Eggleton} P.,  {Mitton} S.,   {Whelan} J.,  eds,
  IAU Symposium Vol. 73, Structure and Evolution of Close Binary Systems. p.~75

\bibitem[\protect\citeauthoryear{Phinney}{Phinney}{2001}]{Phinney2001}
Phinney E.~S.,  2001

\bibitem[\protect\citeauthoryear{Pieroni \& Barausse}{Pieroni \&
  Barausse}{2020}]{Pieroni:2020rob}
Pieroni M.,  Barausse E.,  2020, \mn@doi [JCAP]
  {10.1088/1475-7516/2020/07/021}, 07, 021

\bibitem[\protect\citeauthoryear{Pitkin, Messenger  \& Fan}{Pitkin
  et~al.}{2018}]{PhysRevD.98.063001}
Pitkin M.,  Messenger C.,   Fan X.,  2018, \mn@doi [Phys. Rev. D]
  {10.1103/PhysRevD.98.063001}, 98, 063001

\bibitem[\protect\citeauthoryear{{Portegies Zwart} \& {Verbunt}}{{Portegies
  Zwart} \& {Verbunt}}{1996}]{PZ96}
{Portegies Zwart} S.~F.,  {Verbunt} F.,  1996, \aap, 309, 179

\bibitem[\protect\citeauthoryear{Prince, Tinto, Larson  \& Armstrong}{Prince
  et~al.}{2002}]{aet}
Prince T.~A.,  Tinto M.,  Larson S.~L.,   Armstrong J.~W.,  2002, \mn@doi
  [Phys. Rev. D] {10.1103/PhysRevD.66.122002}, 66, 122002

\bibitem[\protect\citeauthoryear{{Raghavan} et~al.,}{{Raghavan}
  et~al.}{2010}]{Raghavan2010}
{Raghavan} D.,  et~al., 2010, \mn@doi [\apjs] {10.1088/0067-0049/190/1/1},
  \href {https://ui.adsabs.harvard.edu/abs/2010ApJS..190....1R} {190, 1}

\bibitem[\protect\citeauthoryear{{Rebassa-Mansergas}, {Toonen}, {Korol}  \&
  {Torres}}{{Rebassa-Mansergas} et~al.}{2019}]{reb19}
{Rebassa-Mansergas} A.,  {Toonen} S.,  {Korol} V.,   {Torres} S.,  2019,
  \mn@doi [\mnras] {10.1093/mnras/sty2965}, \href
  {https://ui.adsabs.harvard.edu/abs/2019MNRAS.482.3656R} {482, 3656}

\bibitem[\protect\citeauthoryear{{Renzini} \& {Buzzoni}}{{Renzini} \&
  {Buzzoni}}{1986}]{ren86}
{Renzini} A.,  {Buzzoni} A.,  1986, {Spectral Evolution of Galaxies}.
pp 195--231, \mn@doi{10.1007/978-94-009-4598-2_19}

\bibitem[\protect\citeauthoryear{Robson \& Cornish}{Robson \&
  Cornish}{2017}]{Robson2017}
Robson T.,  Cornish N.,  2017, \mn@doi [Class. Quant. Grav.]
  {10.1088/1361-6382/aa9601}, 34, 244002

\bibitem[\protect\citeauthoryear{{Roebber} et~al.,}{{Roebber}
  et~al.}{2020}]{roe20}
{Roebber} E.,  et~al., 2020, \mn@doi [\apjl] {10.3847/2041-8213/ab8ac9}, \href
  {https://ui.adsabs.harvard.edu/abs/2020ApJ...894L..15R} {894, L15}

\bibitem[\protect\citeauthoryear{{Ruiter}, {Belczynski}, {Benacquista}  \&
  {Holley-Bockelmann}}{{Ruiter} et~al.}{2009a}]{Ruiter2009}
{Ruiter} A.~J.,  {Belczynski} K.,  {Benacquista} M.,   {Holley-Bockelmann} K.,
  2009a, \mn@doi [\apj] {10.1088/0004-637X/693/1/383}, \href
  {http://adsabs.harvard.edu/abs/2009ApJ...693..383R} {693, 383}

\bibitem[\protect\citeauthoryear{{Ruiter}, {Belczynski}  \& {Fryer}}{{Ruiter}
  et~al.}{2009b}]{Ruitr2009snIa}
{Ruiter} A.~J.,  {Belczynski} K.,   {Fryer} C.,  2009b, \mn@doi [\apj]
  {10.1088/0004-637X/699/2/2026}, \href
  {https://ui.adsabs.harvard.edu/abs/2009ApJ...699.2026R} {699, 2026}

\bibitem[\protect\citeauthoryear{{Ruiter}, {Belczynski}, {Benacquista},
  {Larson}  \& {Williams}}{{Ruiter} et~al.}{2010}]{rui10}
{Ruiter} A.~J.,  {Belczynski} K.,  {Benacquista} M.,  {Larson} S.~L.,
  {Williams} G.,  2010, \mn@doi [\apj] {10.1088/0004-637X/717/2/1006}, \href
  {https://ui.adsabs.harvard.edu/abs/2010ApJ...717.1006R} {717, 1006}

\bibitem[\protect\citeauthoryear{Seoane et~al.}{Seoane
  et~al.}{2022}]{missionduration}
Seoane P.~A.,  et~al., 2022, \mn@doi [Gen. Rel. Grav.]
  {10.1007/s10714-021-02889-x}, 54, 3

\bibitem[\protect\citeauthoryear{{Sesana}, {Vecchio}  \& {Volonteri}}{{Sesana}
  et~al.}{2009}]{Sesana2009}
{Sesana} A.,  {Vecchio} A.,   {Volonteri} M.,  2009, \mn@doi [\mnras]
  {10.1111/j.1365-2966.2009.14499.x}, \href
  {https://ui.adsabs.harvard.edu/abs/2009MNRAS.394.2255S} {394, 2255}

\bibitem[\protect\citeauthoryear{{Shen}}{{Shen}}{2015}]{Shen2015}
{Shen} K.~J.,  2015, \mn@doi [\apjl] {10.1088/2041-8205/805/1/L6}, \href
  {https://ui.adsabs.harvard.edu/abs/2015ApJ...805L...6S} {805, L6}

\bibitem[\protect\citeauthoryear{{Shen}, {Bildsten}, {Kasen}  \&
  {Quataert}}{{Shen} et~al.}{2012}]{Shen2012}
{Shen} K.~J.,  {Bildsten} L.,  {Kasen} D.,   {Quataert} E.,  2012, \mn@doi
  [\apj] {10.1088/0004-637X/748/1/35}, \href
  {https://ui.adsabs.harvard.edu/abs/2012ApJ...748...35S} {748, 35}

\bibitem[\protect\citeauthoryear{{Sofue}, {Honma}  \& {Omodaka}}{{Sofue}
  et~al.}{2009}]{Sofue2009}
{Sofue} Y.,  {Honma} M.,   {Omodaka} T.,  2009, \pasj, 61, 227

\bibitem[\protect\citeauthoryear{{Tamanini}, {Caprini}, {Barausse}, {Sesana},
  {Klein}  \& {Petiteau}}{{Tamanini} et~al.}{2016}]{Tamanini2016}
{Tamanini} N.,  {Caprini} C.,  {Barausse} E.,  {Sesana} A.,  {Klein} A.,
  {Petiteau} A.,  2016, \mn@doi [\jcap] {10.1088/1475-7516/2016/04/002}, \href
  {http://adsabs.harvard.edu/abs/2016JCAP...04..002T} {4, 002}

\bibitem[\protect\citeauthoryear{Taylor \& Gerosa}{Taylor \&
  Gerosa}{2018}]{PhysRevD.98.083017}
Taylor S.~R.,  Gerosa D.,  2018, \mn@doi [Phys. Rev. D]
  {10.1103/PhysRevD.98.083017}, 98, 083017

\bibitem[\protect\citeauthoryear{{Thiele}, {Breivik}  \& {Sanderson}}{{Thiele}
  et~al.}{2021}]{thi21}
{Thiele} S.,  {Breivik} K.,   {Sanderson} R.~E.,  2021, arXiv e-prints, \href
  {https://ui.adsabs.harvard.edu/abs/2021arXiv211113700T} {p. arXiv:2111.13700}

\bibitem[\protect\citeauthoryear{Timpano, Rubbo  \& Cornish}{Timpano
  et~al.}{2006}]{Timpano2006}
Timpano S.~E.,  Rubbo L.~J.,   Cornish N.~J.,  2006, \mn@doi [Phys. Rev. D]
  {10.1103/PhysRevD.73.122001}, 73, 122001

\bibitem[\protect\citeauthoryear{Tinto \& Dhurandhar}{Tinto \&
  Dhurandhar}{2005}]{tdi}
Tinto M.,  Dhurandhar S.~V.,  2005, \mn@doi [Living Reviews in Relativity]
  {10.12942/lrr-2005-4}, 8, 4

\bibitem[\protect\citeauthoryear{{Toonen}, {Nelemans}  \& {Portegies
  Zwart}}{{Toonen} et~al.}{2012}]{Toonen12}
{Toonen} S.,  {Nelemans} G.,   {Portegies Zwart} S.,  2012, \aap, 546, A70

\bibitem[\protect\citeauthoryear{{Toonen}, {Hollands}, {G{\"a}nsicke}  \&
  {Boekholt}}{{Toonen} et~al.}{2017}]{Toonen2017}
{Toonen} S.,  {Hollands} M.,  {G{\"a}nsicke} B.~T.,   {Boekholt} T.,  2017,
  \mn@doi [\aap] {10.1051/0004-6361/201629978}, \href
  {https://ui.adsabs.harvard.edu/abs/2017A&A...602A..16T} {602, A16}

\bibitem[\protect\citeauthoryear{Vallisneri}{Vallisneri}{2008}]{PhysRevD.77.042001}
Vallisneri M.,  2008, \mn@doi [Phys. Rev. D] {10.1103/PhysRevD.77.042001}, 77,
  042001

\bibitem[\protect\citeauthoryear{{Vigna-G{\'o}mez} et~al.,}{{Vigna-G{\'o}mez}
  et~al.}{2020}]{vig20}
{Vigna-G{\'o}mez} A.,  et~al., 2020, \mn@doi [\pasa] {10.1017/pasa.2020.31},
  \href {https://ui.adsabs.harvard.edu/abs/2020PASA...37...38V} {37, e038}

\bibitem[\protect\citeauthoryear{{Wagg}, {Broekgaarden}, {de Mink}, {van Son},
  {Frankel}  \& {Justham}}{{Wagg} et~al.}{2021}]{wag21}
{Wagg} T.,  {Broekgaarden} F.~S.,  {de Mink} S.~E.,  {van Son} L. A.~C.,
  {Frankel} N.,   {Justham} S.,  2021, arXiv e-prints, \href
  {https://ui.adsabs.harvard.edu/abs/2021arXiv211113704W} {p. arXiv:2111.13704}

\bibitem[\protect\citeauthoryear{{Webbink}}{{Webbink}}{1984}]{Webbink1984}
{Webbink} R.~F.,  1984, \apj, 277, 355

\bibitem[\protect\citeauthoryear{{Wilhelm}, {Korol}, {Rossi}  \&
  {D'Onghia}}{{Wilhelm} et~al.}{2021}]{Wilhelm2021}
{Wilhelm} M. J.~C.,  {Korol} V.,  {Rossi} E.~M.,   {D'Onghia} E.,  2021,
  \mn@doi [\mnras] {10.1093/mnras/staa3457}, \href
  {https://ui.adsabs.harvard.edu/abs/2021MNRAS.500.4958W} {500, 4958}

\bibitem[\protect\citeauthoryear{{Yu} \& {Jeffery}}{{Yu} \&
  {Jeffery}}{2010}]{yu10}
{Yu} S.,  {Jeffery} C.~S.,  2010, \mn@doi [\aap] {10.1051/0004-6361/201014827},
  \href {https://ui.adsabs.harvard.edu/abs/2010A&A...521A..85Y} {521, A85}

\bibitem[\protect\citeauthoryear{{van der Sluys}, {Verbunt}  \& {Pols}}{{van
  der Sluys} et~al.}{2006}]{slu06}
{van der Sluys} M.~V.,  {Verbunt} F.,   {Pols} O.~R.,  2006, \mn@doi [\aap]
  {10.1051/0004-6361:20065066}, \href
  {https://ui.adsabs.harvard.edu/abs/2006A&A...460..209V} {460, 209}

\makeatother
\end{thebibliography}




\appendix
\section{Complementary Figures}

\begin{figure}
 	\includegraphics[width=.9\linewidth]{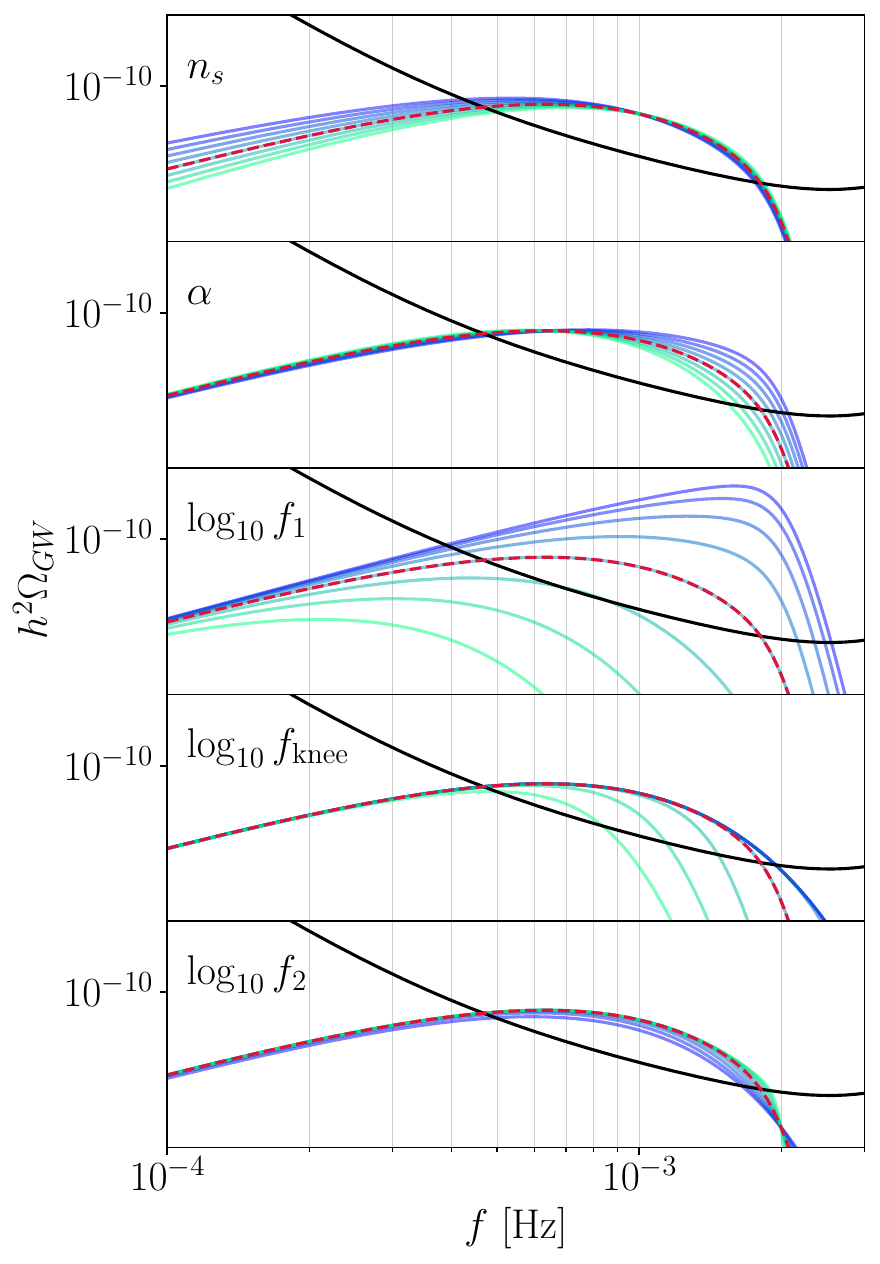}
 	\caption{ \label{fig:model} Understanding the empirical model of~\cref{eq:galfit}. The red dashed line represents the empirical model evaluated at $\theta_\ast = \{ n_s,\, \alpha,\, \log_{10}f_1,\, \log_{10}f_\mathrm{knee},\, \log_{10}f_2\}$, setting their numerical values to $n_s=1.27$, $\alpha=1.31$, $\log_{10}f_1=-3.2$, $\log_{10}f_\mathrm{knee}=-2.7$, and $\log_{10}f_2=-3.5$. The black solid line represents the instrumental noise. At each panel we visualize the impact of the given $i$ parameter $\theta_{\ast,\, i}$ (indicated at the top left of each panel) by keeping the rest o the parameters fixed to the value indicated before. The resulting spectral shapes are given by evaluating~\cref{eq:galfit} at a minimum value of $0.8\times\vec{\theta}_{\ast,\, i}$ up to the maximum of $1.2\times\theta_{\ast,\, i}$, while the color gradient corresponds to the magnitude of the given $\vec{\theta}_{\ast,\, i}$ parameter (green colors correspond to smaller values, as opposed to blue colors). For more details see~\cref{sec:analytic_model} and~\citep{Karnesis2021tsh}.}
 \end{figure}

\begin{figure*}
 	\caption{ \label{fig:zfisher} Histograms of the relative errors of the Galactic binaries waveform parameters for the three cases of simulated data mentioned in~\cref{sec:Shape_Impact}. For the three cases of scale height considered,  $Z_\mathrm{d} = \{100,\, 300,\, 500\}~\mathrm{pc}$, we do not notice significant difference on the parameter estimates uncertainties. This means that the different $Z_\mathrm{d}$ does not affect significantly the resolvability of most of the sources, and therefore the resulting confusion stochastic signal. Thus, it is not possible to make estimates of the $Z_\mathrm{d}$ of the Galactic disk from the stochastic signal of the DWD population alone. See text for more details.}
 	\includegraphics[width=.7\linewidth]{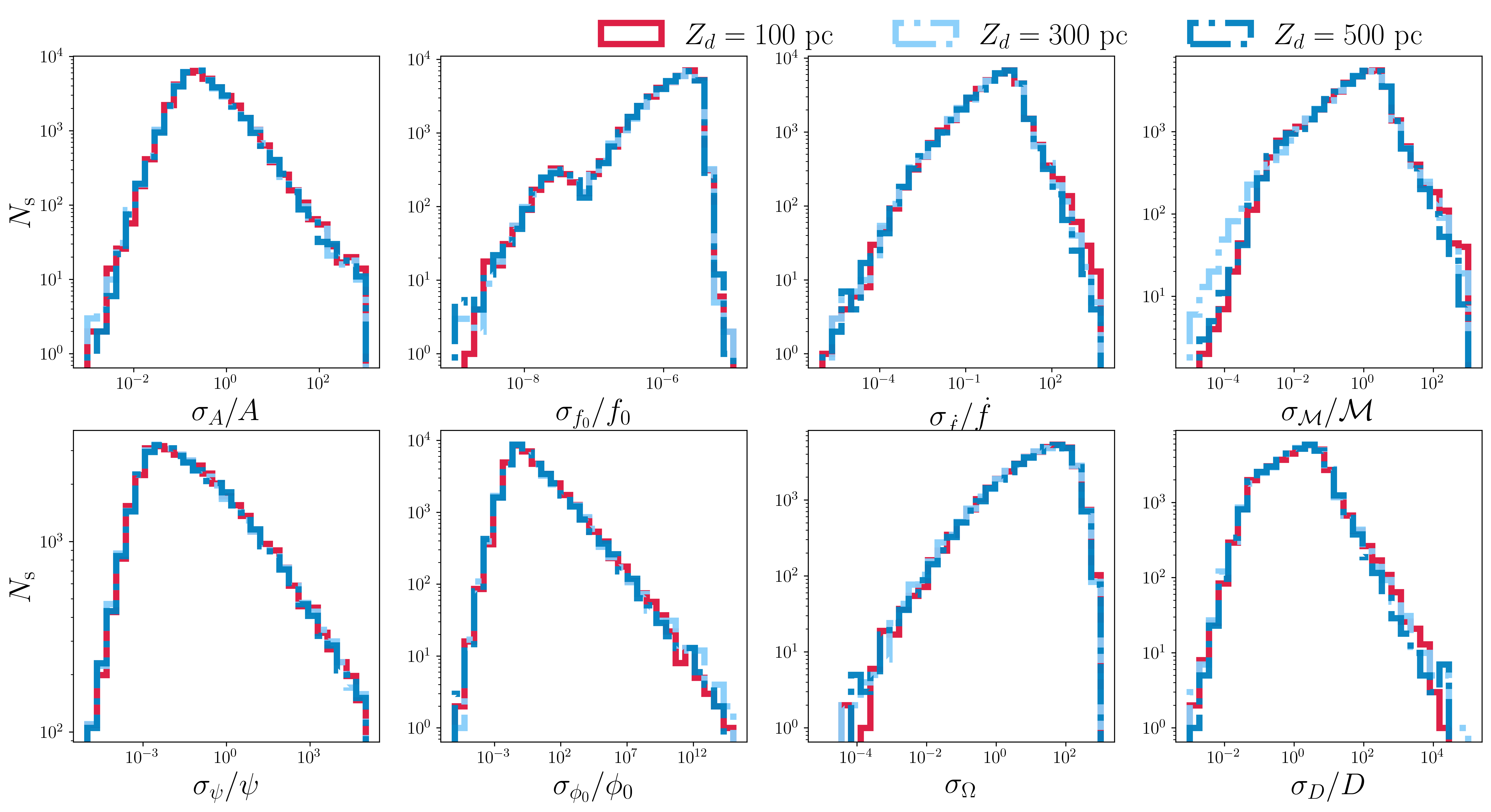}
 \end{figure*}

\begin{figure*}
 	\includegraphics[width=.6\linewidth]{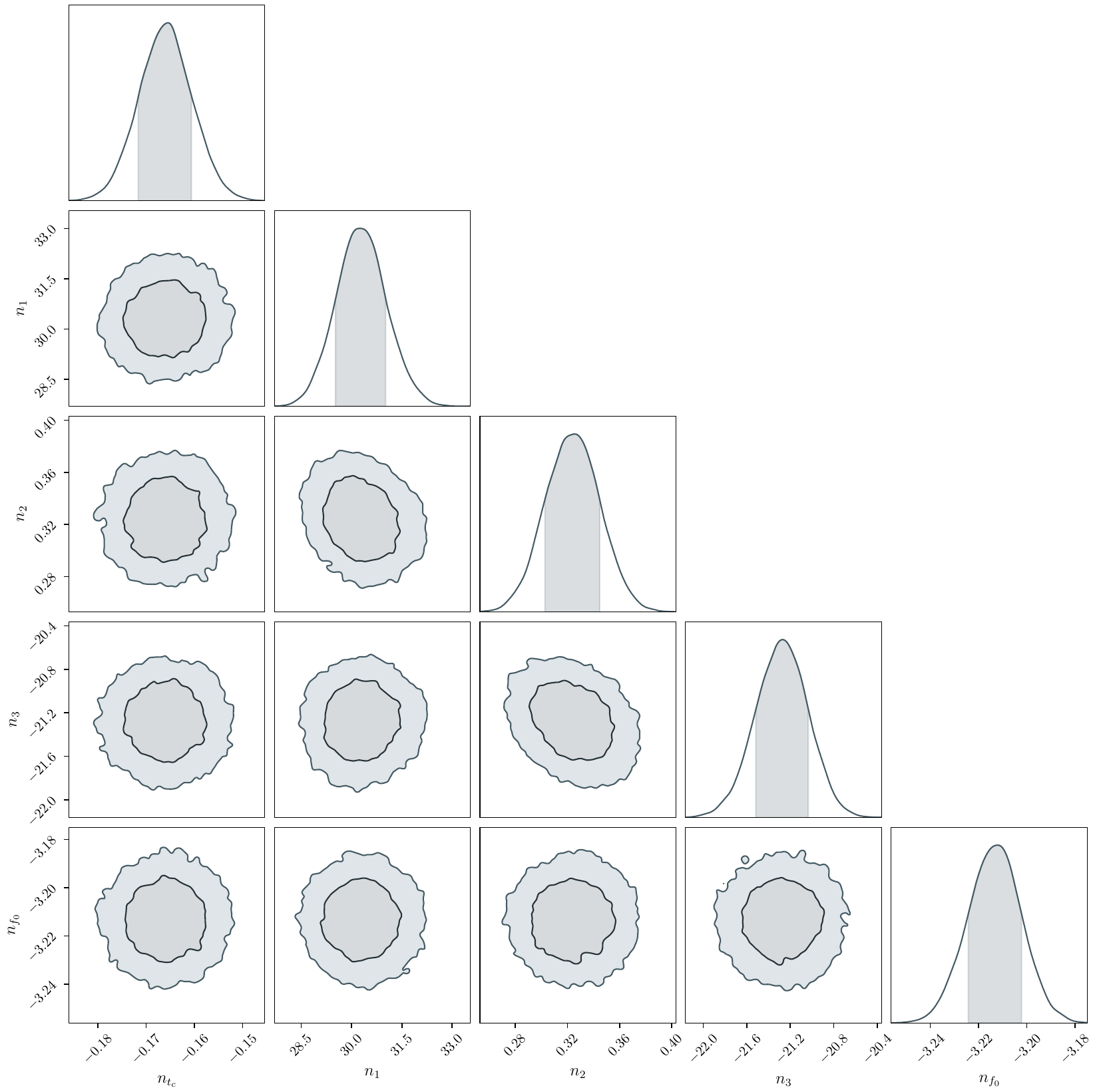}
 	\caption{ \label{fig:posteriors} Posterior densities for the hyper-parameters considered in the hierarchical Bayesian model adopted in~\cref{sec:hierarchical}. The parameters sampled here are basically the shape parameters of hyper-priors for the chirp mass, emission frequency and time of coalescence of each of the resolved source from the data (see text for details).}
 \end{figure*}

In this Appendix section, we provide two figures that are complementary to~\cref{sec:analytic_model}, ~\cref{sec:Shape_Impact}, and~\cref{sec:joint_analysis}. In particular,~\cref{fig:model} demonstrates the impact of the different terms of the analytical model of~\cref{eq:galfit}, while~\cref{fig:zfisher} shows histograms of the relative errors of waveform parameters for the resolved sources, for the three cases of scale height $Z_\mathrm{d}$ considered. In~\cref{fig:posteriors} we show the posterior densities for the hyper-prior parameters considered in the analysis of~\cref{sec:joint_analysis}.

\label{lastpage}
\end{document}